\documentclass[%
 reprint,
 amsmath,amssymb,mathrsfs,
 aps,
]{revtex4-2}

\usepackage{graphicx}
\usepackage[caption=false]{subfig}
\usepackage{dcolumn}
\usepackage{bm}

\usepackage{xcolor}

\usepackage{siunitx}
\usepackage{physics}
\usepackage{multirow}

\usepackage[mathscr]{euscript}
\usepackage{float}
\usepackage{natbib}
\usepackage{braket}

\usepackage[title]{appendix}

\usepackage{color,soul}

\allowdisplaybreaks
\begin{document}
\raggedbottom

\preprint{APS/123-QED}

\title{Axion Dark Matter eXperiment: Run 1A Analysis Details}

\author{C. Boutan}
  \affiliation{
  Pacific Northwest National Laboratory, Richland, WA 99354, USA} 
\author{A. M. Jones}
  \affiliation{
  Pacific Northwest National Laboratory, Richland, WA 99354, USA} 
  \affiliation{currently Open Engineering, Inc., Richland WA 99354, USA}
\author{B. H. LaRoque}
  \affiliation{
  Pacific Northwest National Laboratory, Richland, WA 99354, USA} 
\author{E. Lentz}
 \email[Correspondence to: ]{erik.lentz@pnnl.gov}
  \affiliation{
  Pacific Northwest National Laboratory, Richland, WA 99354, USA} 
\author{N. S. Oblath}
  \affiliation{
  Pacific Northwest National Laboratory, Richland, WA 99354, USA} 
\author{M. S. Taubman}
  \affiliation{
  Pacific Northwest National Laboratory, Richland, WA 99354, USA}
\author{J. Tedeschi}
  \affiliation{
  Pacific Northwest National Laboratory, Richland, WA 99354, USA} 
\author{J. Yang}%
  \affiliation{
  Pacific Northwest National Laboratory, Richland, WA 99354, USA}
  
\author{T. Braine}
  \affiliation{
  University of Washington, Seattle, WA 98195, USA}
\author{N. Crisosto}
  \affiliation{
  University of Washington, Seattle, WA 98195, USA}
\author{S. Kimes}
  \affiliation{
  University of Washington, Seattle, WA 98195, USA}
  \affiliation{currently Microsoft Quantum, Microsoft, Redmond, WA 98052, USA}
\author{R. Ottens}
  \affiliation{
  University of Washington, Seattle, WA 98195, USA}
  \affiliation{currently NASA Goddard Space Flight Center, 8800 Greenbelt Rd, Greenbelt, MD 20771, USA}
\author{L. J Rosenberg}%
  \affiliation{
  University of Washington, Seattle, WA 98195, USA}
\author{G. Rybka}%
  \affiliation{
  University of Washington, Seattle, WA 98195, USA}
\author{D. Will}%
  \affiliation{
  University of Washington, Seattle, WA 98195, USA}
\author{D. Zhang}%
  \affiliation{
  University of Washington, Seattle, WA 98195, USA}

\author{C. Bartram} 
  \affiliation{SLAC National Accelerator Laboratory, Menlo Park, CA 94025, USA}

\author{D. Bowring}
  \affiliation{
  Fermi National Accelerator Laboratory, Batavia IL 60510, USA}
\author{R. Cervantes}
  \affiliation{
  Fermi National Accelerator Laboratory, Batavia IL 60510, USA}
\author{A. S. Chou}
  \affiliation{
  Fermi National Accelerator Laboratory, Batavia IL 60510, USA}
\author{S. Knirck}
  \affiliation{
  Fermi National Accelerator Laboratory, Batavia IL 60510, USA}
\author{D. V. Mitchell}
  \affiliation{
  Fermi National Accelerator Laboratory, Batavia IL 60510, USA}
\author{A. Sonnenschein}
  \affiliation{
  Fermi National Accelerator Laboratory, Batavia IL 60510, USA}
\author{W. Wester}
  \affiliation{
  Fermi National Accelerator Laboratory, Batavia IL 60510, USA}

\author{R. Khatiwada}
  \affiliation{
  Illinois Institute of Technology, Chicago IL 60616, USA}
  \affiliation{
  Fermi National Accelerator Laboratory, Batavia IL 60510, USA}

\author{G. Carosi}
  \affiliation{
  Lawrence Livermore National Laboratory, Livermore, CA 94550, USA}
\author{N. Du}%
  \affiliation{
  Lawrence Livermore National Laboratory, Livermore, CA 94550, USA}
\author{S. Durham}
  \affiliation{
  Lawrence Livermore National Laboratory, Livermore, CA 94550, USA}
\author{S. R. O'Kelley}
   \affiliation{Lawrence Livermore National Laboratory, Livermore, CA 94550, USA}
\author{N. Woollett}
  \affiliation{
  Lawrence Livermore National Laboratory, Livermore, CA 94550, USA}
  \affiliation{currently Rigetti Computing, Oxford, England, UK}

\author{L. D. Duffy}
  \affiliation{
  Los Alamos National Laboratory, Los Alamos, NM 87545, USA}

\author{R. Bradley}
  \affiliation{
  National Radio Astronomy Observatory, Charlottesville, Virginia 22903, USA}

\author{J. Clarke}
  \affiliation{
  University of California, Berkeley, CA 94720, USA}
\author{I. Siddiqi}
  \affiliation{
  University of California, Berkeley, CA 94720, USA}

\author{A. Agrawal}
\affiliation{
University of Chicago, IL 60637, USA}
\affiliation{currently Amazon Web Services Center for Quantum Networking, Boston, MA , USA}
\author{A. V. Dixit}
\affiliation{
University of Chicago, IL 60637, USA}
\affiliation{currently Advanced Microwave Photonics Group, National Institute of Standards and Technology, Gaithersburg, MD 20899, USA}

\author{J. R. Gleason}
  \affiliation{
  University of Florida, Gainesville, FL 32611, USA}
\author{A. T. Hipp }
  \affiliation{
  University of Florida, Gainesville, FL 32611, USA}
\author{S. Jois}
  \affiliation{
  University of Florida, Gainesville, FL 32611, USA}
 \author{P. Sikivie}
  \affiliation{
  University of Florida, Gainesville, FL 32611, USA}
\author{N. S. Sullivan}
  \affiliation{
  University of Florida, Gainesville, FL 32611, USA}
\author{D. B. Tanner}
  \affiliation{
  University of Florida, Gainesville, FL 32611, USA}

\author{J. H. Buckley}
  \affiliation{
  Washington University, St. Louis, MO 63130, USA}
\author{C. Gaikwad}
  \affiliation{
  Washington University, St. Louis, MO 63130, USA}
\author{P. M. Harrington}
  \affiliation{
  Washington University, St. Louis, MO 63130, USA}
  \affiliation{currently Research Laboratory of Electronics, Massachusetts Institute of Technology, Cambridge, MA 02139, USA}
\author{E. A. Henriksen}
  \affiliation{
  Washington University, St. Louis, MO 63130, USA}
\author{J. Hoffman}
  \affiliation{
  Washington University, St. Louis, MO 63130, USA}
\author{K. W. Murch}
  \affiliation{
  Washington University, St. Louis, MO 63130, USA}

\author{E. J. Daw}
  \affiliation{
  University of Sheffield, Sheffield, S10 2TN, UK}
\author{M. G. Perry}
  \affiliation{
  University of Sheffield, Sheffield, S10 2TN, UK}

\author{G. C. Hilton}
  \affiliation{
  National Institute of Standards and Technology, Gaithersburg, MD 20899, USA}

\collaboration{ADMX Collaboration}


\date{\today}

\pagenumbering{gobble}

\begin{abstract}

The ADMX collaboration gathered data for its Run 1A axion dark matter search from January to June 2017, scanning with an axion haloscope over the frequency range $645-680$~MHz ($2.66-2.81 \mu \text{eV}$ in axion mass) at DFSZ sensitivity. The resulting axion search found no axion-like signals comprising all the dark matter in the form of a virialized galactic halo over the entire frequency range, implying lower bound exclusion limits at or below DFSZ coupling at the $90\%$ confidence level. This paper presents expanded details of the axion search analysis of Run 1A, including review of relevant experimental systems, data-taking operations, preparation and interpretation of raw data, axion search methodology, candidate handling, and final axion limits.
    
\end{abstract}

\maketitle{}
\pagenumbering{arabic}

\section{Introduction}

An axion field is a consequence of the Peccei-Quinn solution to the strong-CP problem of particle physics \citep{PQ1977,Weinberg1978,Wilczek1978}. 
The Peccei-Quinn mechanism implies several dynamical processes that produce cold axions in the early Universe. These axions may constitute some fraction or all of the dark matter \citep{ABBOTT1983,PRESKILL1983,DINE1983}. An axion field serves as an excellent candidate for cold dark matter as it has very small primordial velocity dispersion, feeble couplings to itself and to the Standard Model fields.  It describes a very weakly interacting massive particle, the axion, with lifetime much longer than the age of the Universe.
Matter under these conditions is expected to collapse into large scale structures that seed the Universe's galaxies, including our own. For axions to saturate the $\Lambda$CDM model's dark-matter density, many numerical and analytical studies of QCD (quantum chromo-dynamics) prefer an axion mass in the $1-100 \mu \text{eV}$ range \citep{Bonati2016,Berkowitz2015,Borsanyi2016,Ballesteros2017,Dine2017}.

The predicted coupling between axions and photons is model-dependent; in general, axions with dominant hadronic couplings as in the KSVZ (Kim-Shifman-Vainshtein-Zakharov) model \citep{Kim1979,SHIFMAN1980} are predicted to have an axion-photon coupling roughly 2.7 times larger than that of the DFSZ (Dine-Fischler-Srednicki-Zhitnitsky) model \citep{Zhitnitsky1980,DINE1981}. Because the axion-photon coupling is expected to be very small, $O(10^{-17} - 10^{-12})$\,GeV$^{-1}$ over the expected axion mass range, these predicted particles are dubbed \textit{invisible axions}.

\begin{figure}
\centering
\includegraphics[width=\linewidth]{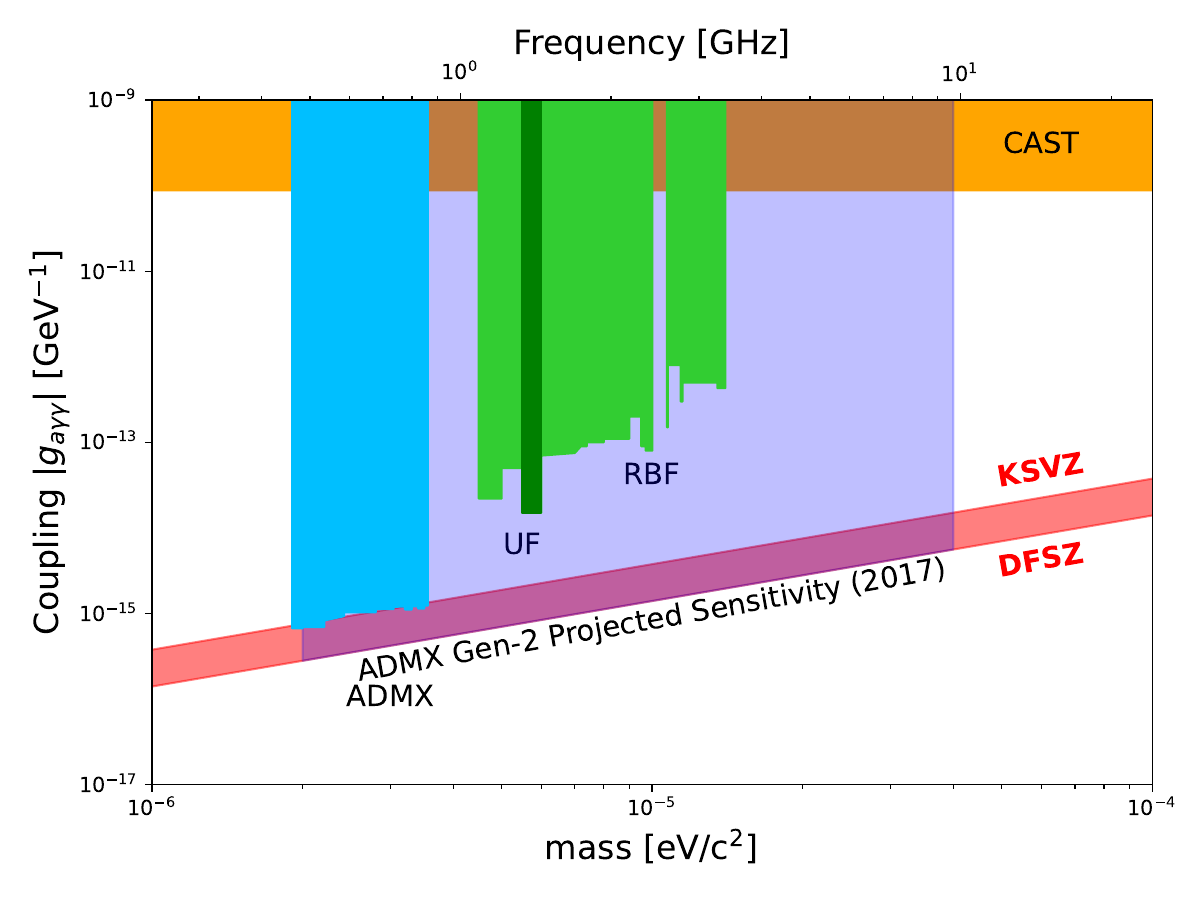}
\caption{Mass and photon coupling parameter space of axion-like particles over the QCD axion dark-matter region. This plot is current as of early 2017, prior to publication of the Run 1A limits \citep{Du2018}. Limits produced on the axion-photon coupling are opaque \citep{RBFLimits1987,Hagmann1990,ASZTALOS201139}, and the ADMX generation 2 discovery potential as projected in early 2017 is displayed in translucent blue \citep{Stern2016}. Benchmark couplings from the KSVZ and DFSZ models are given by the diagonal red band's upper and lower edges respectively. Previous ADMX limits and sensitivity projections assume a local axion dark-matter density of $\rho_a = 0.45$~GeV/cc.}
\label{fig:param_space}
\end{figure}

Direct searches for axions use several techniques, based in either pure laboratory methods or by providing the requisite axions through external sources. The Axion Dark Matter eXperiment (ADMX) utilizes the axion haloscope technique~\citep{Sikivie1983}, and consists of a cold microwave cavity threaded by a static magnetic field coupled to the ambient axion field via an inverse-Primakoff process. Axions are resonantly converted to photons in a cavity mode tuned to the axion field's frequencies, which are centered at $f = \bar{\epsilon} /h$ where $h$ is the Planck constant and $ \bar{\epsilon}$ is the expected total axion energy consisting of the rest mass energy and kinetic energy. The microwave haloscope has proven to be the most sensitive technique to search for axions as the dark matter over the favored range. Axion searches prior to the year 2017, including ADMX, have managed to approach KSVZ sensitivity, but not the stricter DFSZ model under the conservative assumption of a virialized galactic halo. ADMX entered as a DOE Generation 2 dark-matter search in 2014 with the explicit target of reaching DFSZ sensitivity over the frequency range $0.5-10$\,GHz ($2-40 \mu \text{eV}$) \citep{EarthSky2014}, see Fig.~\ref{fig:param_space}, and more recently has been split into G-2 and EFR (extended frequency range) efforts. The first of these searches, referred to as ``Run 1A,'' gathered data between January and June 2017 over the frequency range $645-680$\,MHz ($2.66-2.81 \mu \text{eV}$) and found no axions in that range \citep{Du2018}. This paper explains in more detail the operations and analyses that resulted in the Run 1A search reaching DFSZ limits.

The remainder of the this paper is structured as follows. Section~\ref{sec:Apparatus} provides a brief overview of the ADMX apparatus, with focus on the axion converter-detector assembly, including the cavity and magnet, the radio frequency (RF) receiver chain, and the data acquisition system.
Section~\ref{sec:Operations} presents the data-taking operations, including search guidelines, data-taking cadence, scanning operations, logging of data, and candidate rescans.
Section~\ref{sec:Preparations} reviews the preparations performed on the raw experimental data prior to searching for axion signals, paying particular attention to noise characterization, backgrounds in the integrated power spectra, and the classification of unfit data.
Section~\ref{sec:Statistics} presents the statistical models used to search for the axion, how multiple observations are composed to form the grand statistical spectra, and the process of converting sensitivity spectra to limits.
Section~\ref{sec:Signal} reviews the different axion dark-matter models searched for, including the model used during data-taking operations and the more sophisticated models of the final analysis.
Section~\ref{sec:Candidates} presents the procedure for handling axion candidates during Run 1A, including initial identification using the live analysis output, generation of synthetic axions, rescans and further procedures, residual candidates, and the reintroduction of rescan data for the construction of the grand spectra.
Section~\ref{sec:Limits} consolidates the cumulative findings into a grand spectrum, providing the net Run 1A axion limits.
Section~\ref{sec:Conclusions} summarizes the findings and sets the stage for Run 1B \citep{Braine2020,Bartram2021}.

\section{The Haloscope Apparatus}
\label{sec:Apparatus}

This section provides a summary of the ADMX haloscope apparatus, specifically the magnet, resonant cavity, cold receiver chain, warm receiver, and data acquisition (DAQ) systems. The focus is on components directly relevant for the axion search. These and other supporting components are discussed in more detail in the recent ADMX instrumentation paper of Run 1A and 1B \citep{Khatiwada2020}. 

\begin{figure}[t]  
\includegraphics[width=\linewidth]{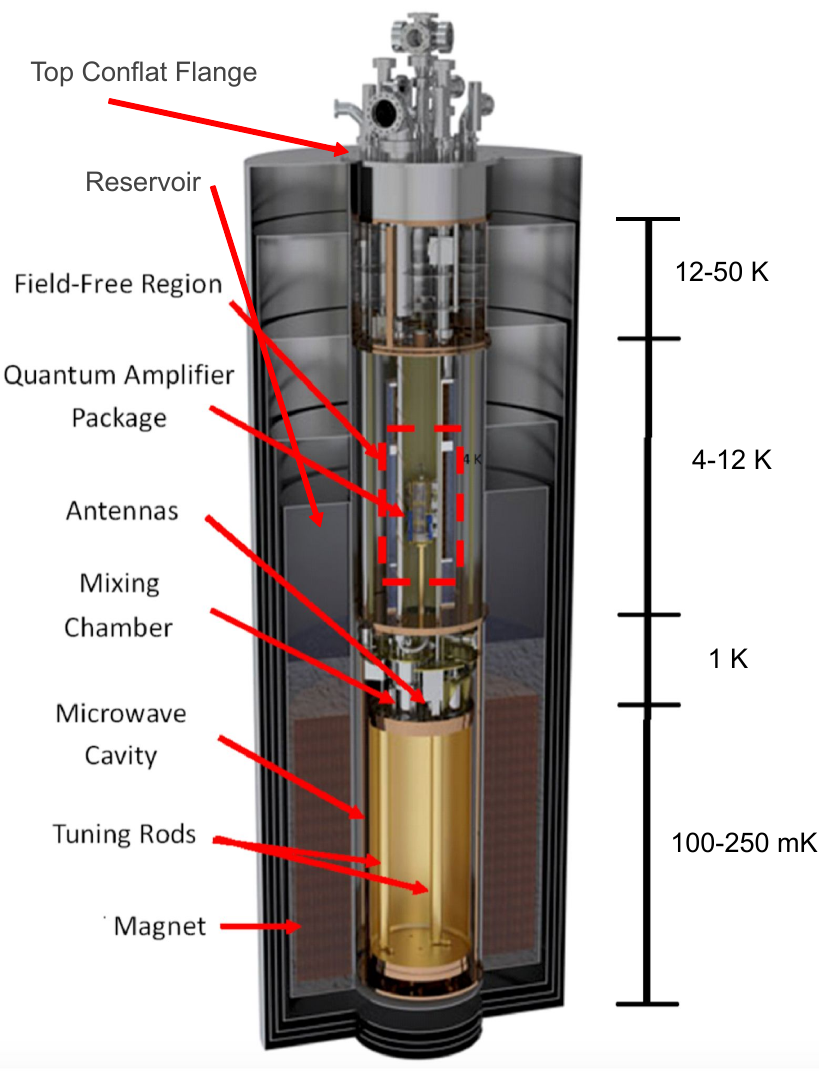}
\centering
\caption{Cutaway model of the ADMX apparatus. The ``Insert'' (below and including the top conflat flange) is comprised of three main sections: ``the cold space'' (below 4 Kelvin),  the ''reservoir'' (4-12 Kelvin), and a 12-50 Kelvin space. The cold space is primarily cooled by a DR below 1 Kelvin (K) and also contains the cavity and radiation shields to prevent components from being radiatively coupled. The ``field-free region'' containing cold electronics components sensitive to high magnetic fields is also in the cold space as it is thermally sunk to the DR. Sitting on top of the cold space is the helium reservoir. This portion is maintained at 4~K and houses a helium reservoir and other components that are thermally sunk to it. Above that is the warm space. This includes baffles, supporting structural components and a top conflat flange, above which are the only components visible during operations. The Insert sits inside a cryostat, which holds the superconducting magnet. The main magnet can generate fields exceeding 8~Tesla in the cavity. Labeled components will be elaborated on later in Section~\ref{sec:Apparatus}. This figure is modified from a version that first appeared in \cite{Khatiwada2020}.}
\label{fig:cutaway}
\end{figure}

The main ADMX apparatus is an Earth-bound tunable haloscope, designed to detect relic axions via microwave power emitted from a resonantly enhanced inverse-Primakoff process. The experimental apparatus, shown in Fig.~\ref{fig:cutaway}, is divided into a graduated cold space, which at its coldest level ($\sim 150$~mK) is regulated by a dilution refrigerator (DR), and surrounded by increasingly warmer spaces leading up to room temperature. The cold space lies primarily within the main magnet housing. The outer steel housings of the magnet cryostat act as shielding against heat and radio frequency interference (RFI) leakage \citep{Yu2004,lyapustin2015dis}. Inside the magnet bore housing is the detector insert, containing the cavity, cold electronics, and sub-Kelvin cryogenics. Two more layers of heat/RFI shields separate the coldest elements, including the resonant cavity, from the $4.2$~K main magnet. Power from the cavity is transmitted through a tunable antenna in the top plate and out of the cold space through an ultra-low noise transmission and amplification chain. This paper concentrates only on the receiver chain used in Run 1A, which tracked a TM$_{010}$-like mode in the cylindrical cavity. In the warm space, the transmitted signal is then mixed down in frequency and digitized as a voltage time series and a frequency power spectrum over a $25$~kHz bandwidth, which roughly matches the bandwidth of the TM$_{010}$-like resonance. These digitized power spectra are where the axion signatures would be expected to appear.

The warm space also includes all equipment outside the magnet shield, and applies to the majority of the cryogenics and gas handling infrastructure. Also at room temperature is the DAQ infrastructure that monitors and controls the components of the warm and cold spaces. The DAQ monitors and controls many physical components of the experiment, such as running the DR, operation of the quantum-limited amplifiers and mechanical control of the cavity, monitoring of the cryogenics, cavity, magnets, and a complex of sensors for magnetic field strength, pressures, cryogen levels and temperatures at every stage of the insert. The lowest layer of the DAQ software is based on EPICS \citep{Khatiwada2020,EPICS_homepage}, which provides a uniform software interface for interaction with the instruments. The DAQ infrastructure also executes the numerous routines used in data taking and monitors its status using a live analysis.

\subsection{Resonant Axion-to-microwave Converter Assembly}

This sub-section reviews the primary components responsible for converting ambient axions into microwave photons: the main magnet and cavity.

\subsubsection{Main Magnet}

\begin{figure}[H]  
\includegraphics[width=\linewidth]{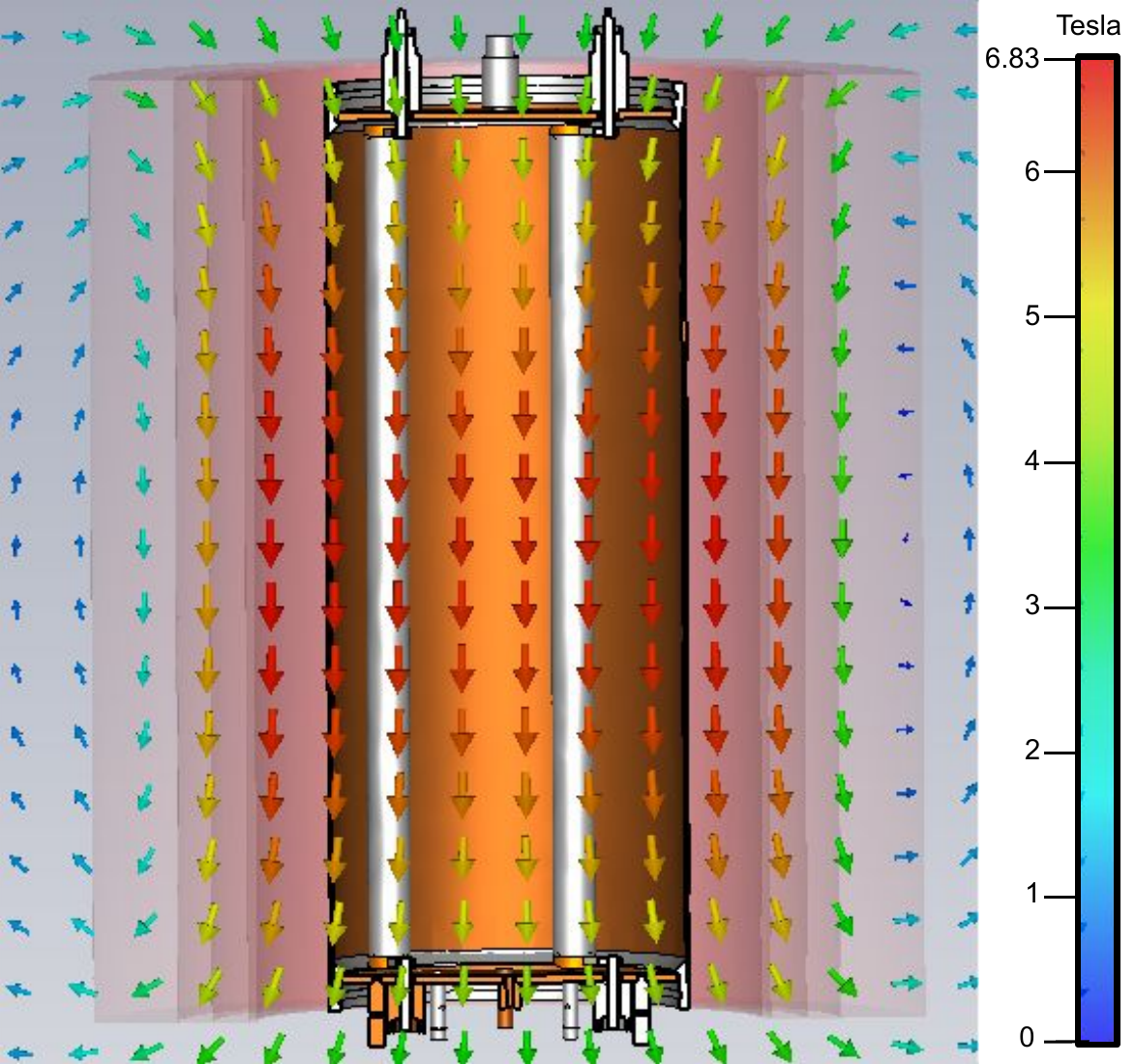}
\centering
\caption{Cutaway model of the ADMX main magnet's field, simulated in the absence of the rest of the apparatus using the COMSOL multi-physics simulation software. The cutaway model of the cavity is included for reference. The size of the arrow indicates the strength of the field at that point in addition to the color scale.}
\label{fig:magnet}
\end{figure}

The static magnetic field is supplied by a superconducting magnet manufactured by Wang NMR of Livermore, CA \cite{Yu2004,Hotz2013dis}. The magnet consists of niobium-titanium windings immersed in liquid $^4$He. During normal Run 1A data-taking operations, the field at the center of the solenoid was $6.8$~Tesla, falling off to about $70\%$ of this value at the center of the end plates of the cavity, as seen in Fig.~\ref{fig:magnet}.
The superconducting coil is a $1.12$~meter tall solenoid with a $60$~cm inner diameter bore. The magnet winding is composed of four concentric superconducting solenoids, containing $99$~km of copper-stabilized niobium-titanium wire wound around a stainless steel spool piece and potted in epoxy. The rated maximum field at the center of the magnet is $8.5$~Tesla at a current of $248.96$~Amperes and a stored energy of $16.54$~MJoules. The central field of the main magnet runs linearly with the current through the coils
\begin{equation}
    |B|_{max} = 8.5 \text{~Tesla} \left(\frac{ I }{248.96 \text{Ampere}}\right), \label{magnetBmax}
\end{equation}
where $I$ is the supplied current to the coils. The main magnet’s current is continuously maintained by a power supply. Once cooled to operating temperature, the magnet requires a supply of approximately $2000$~liters of liquid helium per month for continuous cooling during data-taking operations.

\subsubsection{Cavity}

Axions passing through the field of the main magnet transition into microwave photons with some small probability. To capture and maximize this conversion, a highly conductive cavity is placed in the magnet bore at the center of the solenoid field.
The ADMX Run 1A cavity is a $136$~liter copper-plated stainless-steel cylinder with two copper tuning rods. The cylinder is made of a tube-section forming the cavity walls plus two removable end caps. The end caps form a low-resistance connection to the walls via a knife edge on the walls pressed firmly into the end plates. Both the stainless-steel cavity and the copper tuning rods are plated with OFHC (oxygen free high conductivity) copper with a minimum thickness of $0.075$~mm and then annealed for 8 hours at 400~Celsius in vacuum. The annealing process further increases the grain-size of the copper crystals allowing for longer electron scattering path lengths as the copper is cooled and enters the anomalous skin depth regime. The quality factor $Q$ of the tracked TM$_{010}$-like mode in the unloaded cavity ranges from $20,000$ at room temperature to $160,000$ at cryogenic temperatures \citep{Hotz2013dis}. The unloaded Q of the cavity primarily depends on the resistive skin losses of the copper plating. The oscillating electric fields of the cavity mode penetrate the resistive copper walls of the cavity. For more information on the construction of the Run 1A cavity, see \citet{Hotz2013dis,Khatiwada2020}.

The cumulative interaction strength between the ambient axions and electromagnetic fields over the interior of the cavity is given in natural units by the Lagrangian 
\begin{equation}
    L_{\text{cav}} = - \int_V d^3 x \frac{\alpha g_{ \gamma}}{\pi f_a} a \vec{E} \cdot \vec{B},
\end{equation}
where $V$ is the inner volume of the cavity, $\alpha$ is the fine structure constant, $g_{\gamma}$ is the dimension-less model-dependent coupling constant, $f_a$ is the Peccei-Quinn symmetry-breaking scale, $a$ is the axion field, $\vec{E}$ is the electric field strength, and $\vec{B}$ is the magnetic field strength. 
For a highly resonant cavity threaded by a strong magnetic field, passing relic axions may be seen to act as an external driving force on cavity wave-forms with non-trivial Lagrangian $L_{\text{cav}}$ so long as the axion-photon coupling is feeble enough that the probability of being converted is small. The dynamics of the system become resonant around the cavity modes as the boundary fields and losses per cycle are small in comparison to the stored energy.

The power deposited into the cavity from a monotone source averaged over times much longer than the coherence time of the cavity mode, $t_{coh} = Q_L/\nu_0$, where $\nu_0$ is the mode center frequency and $Q_L$ is the loaded quality factor, in SI units is found to be 
\begin{widetext}
\begin{align}
    \left< P_{cav} \right>(\nu) &= \frac{\epsilon_0 \alpha^2 c^2}{\pi^2 f_a^2} g_{\gamma}^2 V |B|_{max}^2 C \nu Q_L \left< |a|^2 \right>(\nu) T_{\nu_0} (\nu)  \nonumber \\
    &= 1.9 \times 10^{-23} W \left(\frac{g_{\gamma}}{0.97}\right)^2 \left(\frac{V}{136~\text{l}}\right) \left(\frac{B_{max}}{6.8~\text{T}}\right)^2 \left(\frac{C_{\text{mode}}}{0.4}\right) \left(\frac{\nu}{650~\text{MHz}}\right) \left(\frac{Q_L}{50,000}\right) \left(\frac{\rho}{0.45~\text{GeV cm}^{-3}}\right) T_{\nu_0} (\nu) \label{eqn:axionpwr}.
\end{align}
\end{widetext}
Here $\epsilon_0$ is the permittivity of free space, $c$ is the speed of light in vacuum,  $C_{\text{mode}}$ is the form factor quantifying the mode and magnetic field alignment, and $T_{\nu_0} (\nu)$ is the mode envelope shape, which is expected to follow a Lorentzian form
\begin{equation}
    T_{\nu_0} (\nu) = \frac{1}{1 + 4 Q^2 \frac{(\nu-\nu_0)^2}{\nu_0^2}}. \label{eqn:Lorentzian}
\end{equation}
The cavity form factor parametrizes the overlap of the magnetic field and electric field in the cavity 
\begin{equation}
    C = \frac{ \left( \int_V d^3x \vec{E} \cdot \vec{B} \right)^2 }{ \left(\int_V d^3x |\vec{E}|^2 \right) |B_{max}|^2 V }.
\end{equation}
The magnetic field under is dominated by several orders of magnitude by the external main magnet under normal operating conditions, which has profile $\vec{B}_0$. The electric field in the vicinity of a single cavity mode is dominated by the mode's form $\vec{E}_{\xi}$ where $\xi$ is the multi-index of the mode wave-form, which will often be parameterized by modified cylindrical harmonics $(n,l,m,X)$ where the last index $X$ will be used to distinguish modes split by the axial-symmetry-breaking tuning rods. The form factor of the $\xi$ mode is then well approximated by
\begin{equation}
    C_{\xi} \approx \frac{ \left( \int_V d^3x \vec{E}_{\xi} \cdot \vec{B}_0 \right)^2 }{ \left(\int_V d^3x |\vec{E}_{\xi}|^2 \right) |B_{max}|^2 V }. \label{eqn:modeformfactor}
\end{equation}

The main magnet's field is oriented vertically along the cavity's axis, though it does diverge some at either end, see Fig.~\ref{fig:magnet}, making the value of the form factor primarily dependent on the shape of the mode's axial electric field. For the ADMX cavity, there are, among others, transverse electric (TE) modes and transverse magnetic (TM) modes. The TM modes are the only category that contain an axial electric field desired for large form factors. The axial electric field for a TM$_{nlm}$ mode of an empty right-circular cylinder is
\begin{equation}
    \vec{E}_{nlm}(t,\rho, \phi, z) = \hat{z} E_{amp}(t) J_m(x_{ml} \rho / R) e^{\pm i m \phi} \cos \left( \frac{n \pi z}{d} \right)
\end{equation}
where $E_{amp}(t)$ is the time dependent component of the field, $J_m$ is a cylindrical Bessel function, $x_{ml}$ is the $l$-th root of $J_m(x) = 0$, $R$ is the cavity radius, and $d$ is the cavity height. For the rod-less cavity and magnet configuration used in ADMX, the chosen TM$_{010}$-like mode maximizes the form factor $C_{m}$.

Two copper-plated stainless steel tuning rods are placed in the cavity in order to tune the axion-coupled modes. The rods run through the length of the cavity and are mounted on rotating alumina oxide rotary armatures. The rods are $0.05$~m in diameter \cite{Khatiwada2020,Hotz2013dis}. The rods ideally create null boundary conditions in the electric field of the cavity mode, effectively shrinking the extent of the cavity in the horizontal plane, increasing the TM mode frequencies, and splitting modes by breaking the axial symmetry. The presence of the tuning rods also reduce the quality factor from that expected of an empty cavity by about a factor of two. Identifying the split TM$_{010}$-like modes can be done heuristically or through simulation by increasing the thickness of the rods from zero (bare cavity) to their physical dimensions \citep{Boutan:2017oxg}. The TM$_{010}$-like mode branch tracked in Run 1A is identified with the name TM$_{010c}$ in \citet{lyapustin2015dis}. Rotating the rods from near the wall of the cavity to near the center tunes the TM$_{010}$-like mode from roughly $580$~MHz to $890$~MHz. The rods are moved by stepper motors located on top of the experimental insert. The stepper motors are connected to long G10 shafts that drive gearboxes on top of the cavity, which in turn are connected to the alumina shafts of the tuning rods. The gear boxes consist of two anti-backlash worm gear reductions, both geared down by 140:1 for a total reduction of 19600:1. Combined with the precision of the stepper motors, the angle of the rods can be stepped at the level of micro-radians. In practice, this stepping resolution allowed for tuning the cavity at 100-200~Hz per step during Run 1A operations.

\begin{figure}
\centering
\includegraphics[width=\linewidth]{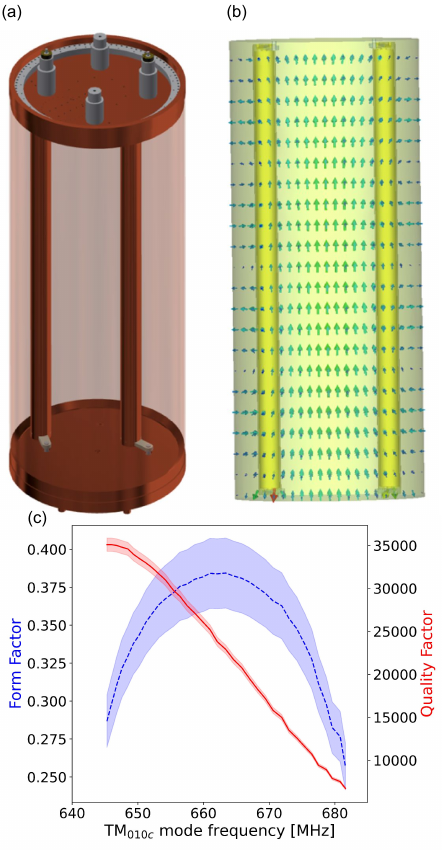}
\caption{ADMX main cavity and TM$_{010}$-like simulated mode properties: (a) view of the interior of the main cavity; (b) TM$_{010}$-like mode electric field in side view; (c) simulated mode form factor (blue dashed) and measured cyogenic loaded quality factor (red solid) over the Run 1A tuning path. Errors (shaded regions) in the form factor and quality factor were computed to be 6\% and 2.2\% respectively, as discussed in Section~\ref{sec:Errors} and presented in Table~\ref{table:errors}.}
\label{fig:TM010}
\end{figure}

The motion of the rods is most effective in tuning the TM modes, with TE frequencies remaining nearly constant. However, when the TM$_{010}$-like mode crosses another mode, their waveforms mix and share energy. This mixing degrades both the form factor $C_{nlmX}$ and quality factor and severely complicates the flow of power. As such, the axion search becomes insensitive at these crossings. The two-rod system can be maneuvered to avoid many of these crossings, but not all. For Run 1A, one of the rods was put in the wall position while the other was left free to tune. Frequencies of the cold space circulators and the MSA (microstrip SQUID amplifier) coincided with a range of good form factor between two major mode crossings, as seen in Figs.~\ref{fig:TM010},~\ref{fig:rod_tuning}. The form factor is simulated over this range using FEM (finite element method) multi-physics software Comsol \citep{Khatiwada2020}. 

The cavity contains three microwave ports: one ``weak'' port on the bottom plate and two tunable ``major'' ports on the top plate. The ports allow for the extraction or insertion of RF power with the cavity. The weak port is a fixed, short antenna that is extremely under-coupled to the TM$_{010}$-like mode. The weak port can be used to inject signals into the cavity without degrading the loaded quality factor of the mode $Q_L$ or extracting signal power. The major ports consist of movable antennae that can be inserted or withdrawn via stepper motors attached to linear gear drives to vary the coupling strength. Like the tuning rods, the major ports' linear drives are actuated by stepper motors attached to long G10 shafts which turn 140:1 anti-backlash worm gear drives. The intention of having two major ports in Run 1A was to perform two simultaneous axion searches on two TM modes, a TM$_{010}$-like and a TM$_{020}$-like. However, for reasons covered in \citet{Khatiwada2020}, only the TM$_{010}$-like mode was observed and the other port was uncoupled from the cavity. Only the TM$_{010}$ port and the connected receiver chain will be referenced from here on.

\begin{figure}
\centering
\includegraphics[width=\linewidth]{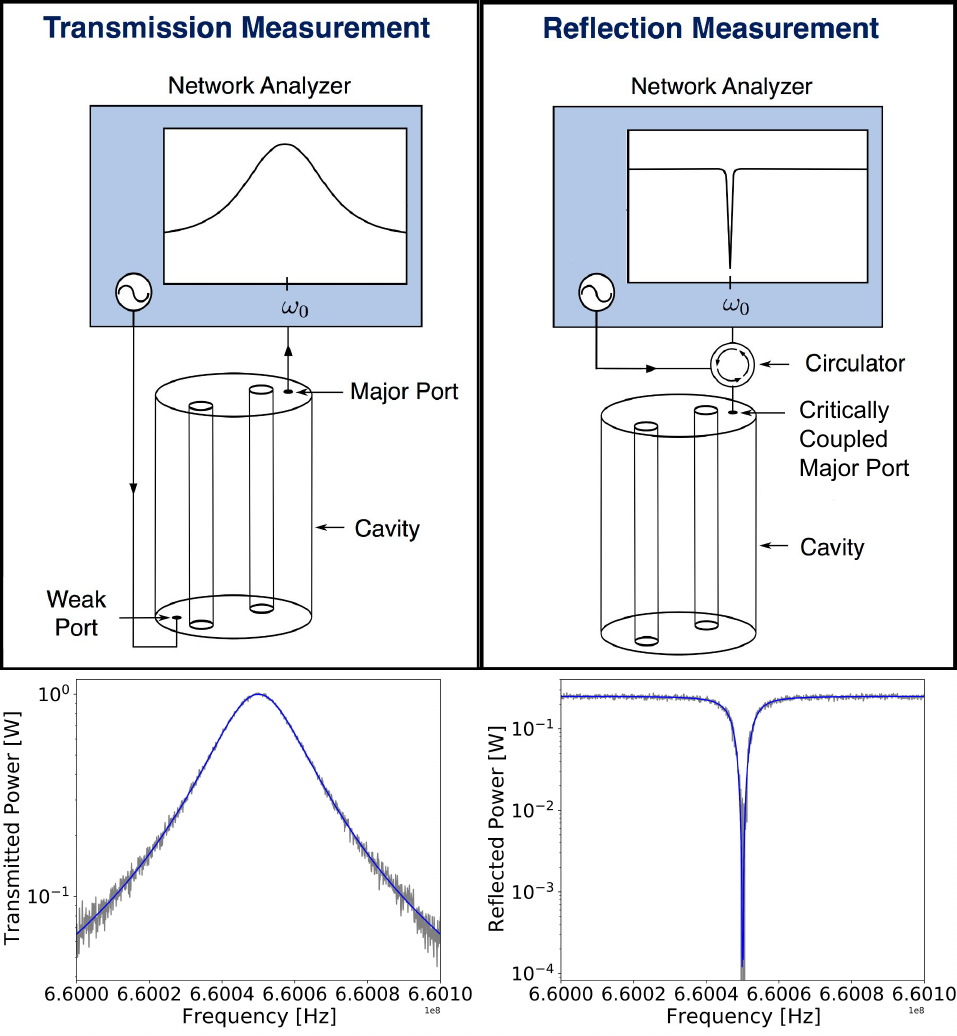}
\caption{Sample transfer functions of the resonant cavity around well-coupled mode (TM$_{010}$-like). The mode is centered at 660.05~MHz. (Left) (top) Illustration of the mode transmission function S$_{12}$ from the weak port through the main port with a Lorentzian envelope about the mode center (Eqn.~\ref{eqn:Lorentzian}). (bottom) Example S$_{12}$ scan with fit Lortentzian with the cavity at sub-Kelvin temperature. (Right) (top) Illustration of the mode reflection function S$_{11}$ in/out of the main port with envelope of the form in Eqn.~\ref{eqn:cavity_reflection}. (bottom) Example S$_{11}$ scan with the best fit envelope with cavity at sub-Kelvin temperature. }
\label{fig:mode_shape}
\end{figure}

A vector network analyzer (VNA) can be used to measure the TM$_{010}$-like mode's frequency, quality factor, and the major port's coupling to the cavity. The VNA measures the swept transfer function (S$_{12}$) of the cavity by injecting a tone into the weak port of the cavity and measuring the same tone’s amplitude transmitted through the major port. The frequency of the injected tone is swept across the band of the resonant mode. At frequencies far outside the central resonant frequency of the cavity, the injected tone is largely reflected by the cavity at the weak port or absorbed by the cavity walls so almost none is extracted at the major port. On resonance, the injected tone enters the cavity, excites the mode, and its power is extracted at the major port with far less loss. The output of the VNA is the swept response across the cavity mode. The expected response of an unmixed isolated mode is the Lorentzian distribution of Eqn.~\ref{eqn:Lorentzian}, from which the central frequency and Q-width can be extracted from a fit, seen in Fig.~\ref{fig:mode_shape}.

The VNA also determines the coupling properties between the major port and cavity via a swept reflection measurement (S$_{11}$). The VNA sends power towards the major port of the cavity through a circulator. A circulator allows for the directional injection of power along a transmission line. The power incident on the cavity reflects back with the form 
\begin{equation}
    \Gamma_{\nu_o}(\nu) = \frac{\beta - 1 + Q_L^2 \left(\frac{\nu - \nu_o}{\nu_o} \right)^2 - 2 i \beta Q_L \frac{\nu - \nu_o}{\nu_o}}{1 + 4 Q_L^2 \left(\frac{\nu - \nu_o}{\nu_o} \right)^2}, \label{eqn:cavity_reflection} 
\end{equation}
where $\beta$ is the coupling strength parameter. The coupling strength can be expressed as
\begin{equation}
    \beta = \frac{Q_{0}}{Q_{\text{ext}}},
\end{equation}
where $Q_{0}$ is the cavity $Q$-factor with the major port uncoupled, and $Q_{\text{ext}}$ is the contribution to the quality factor from external losses such as the major port. For a given coupling, the response is total reflection off resonance and a dip on resonance where power is absorbed by the coupled cavity. Critical coupling occurs when $\beta = 1$ and all power passes into the cavity at the central frequency. A good impedance match is marked by a deep trough in the reflected baseline on resonance. The depth of an antenna is adjusted to maximize trough depth for the TM$_{010}$-like mode as proxy for critical coupling. More precise analyses of coupling strength that can track overcoupling and other conditions have been implemented in subsequent runs \cite{Bartram2021}. Conventionally, when the difference between the minima of the trough and the off-resonance baseline reaches $-30$~dB the antenna is considered critically coupled. Such a trough means that only 0.1 percent of the on-resonance incident power is reflected. External losses from the major port lower the mode quality factor as
\begin{equation}
    \frac{1}{Q_{\text{L}}} = \frac{1}{Q_{0}} + \frac{1}{Q_{\text{ext}}},
\end{equation}
where $Q_{\text{L}}$ is the quality factor of the loaded cavity. During critical coupling, half of the power is lost to the major port, meaning $Q_{\text{loaded}} = Q_{\text{free}}/2$. Run 1A is configured to run at critical coupling, meaning at best only half of the axion power generated in the cavity is expected to leave through the major port.

\subsection{Run 1A Cold Receiver Chain}

\begin{figure}
\centering
\includegraphics[width=\linewidth]{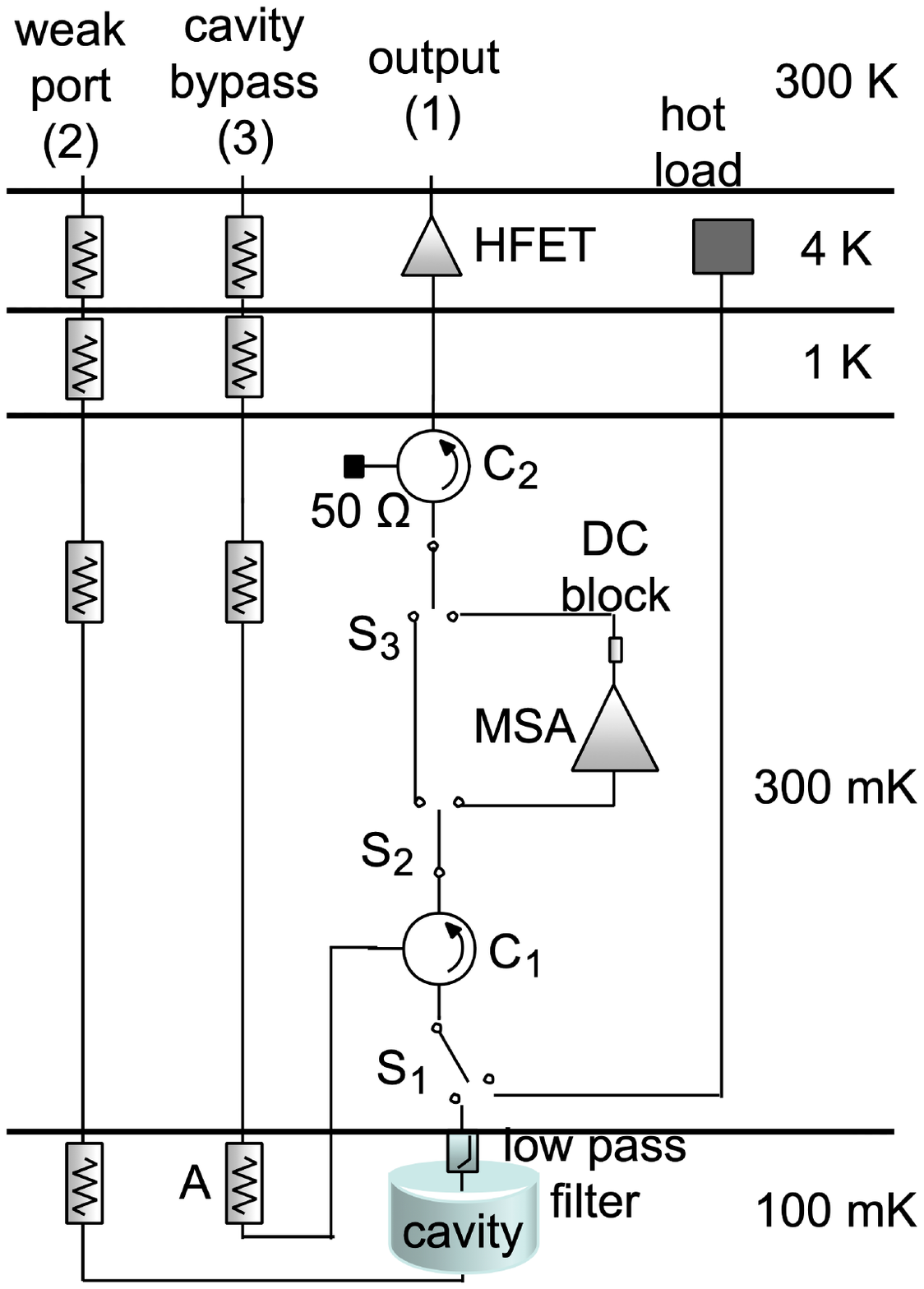}
\caption{Schematic of cold RF receiver for Run
1A, first shown in \citet{Khatiwada2020}. Labeled components include circulators C$_1$ and C$_2$ to guide cavity bypass and reflection measurements, a low pass filter to limit incident thermal radiation at higher frequencies to be emitted by the receiver, switches (S$_{*}$), MSA, DC block,
variable resistor (hot load), and the HFET (heterostructure field-effect transistor) amplifier. All components are connected with coaxial cabling. Switches allow the characterization of the system noise by toggling between a heated ``hot load'' and the main cavity (S$_{1}$), and bypass the MSA (S$_{2}$ and S$_{3}$). Other pathways are for auxilliary measurements such as mode mapping (2) and MSA tuning (3), and are not used for axion observation.}
\label{fig:rec_chain}
\end{figure}

The TM$_{010}$ major port connects to the RF receiver chain, as seen in Fig.~\ref{fig:rec_chain}. The chain begins with a low-pass filter and runs through the cold space where the quantum-limited electronics are housed. A bucking magnet surrounds the quantum amplifiers and other electronics sensitive to stray magnetic fields, actively canceling the field from the main magnet to tens of Gauss. Two hall probes are located in the field free region of bucking coil to confirm the field is cancelled to within a few Gauss \citep{Khatiwada2020}. The electronics inside the field-free region are contained in an OFHC copper frame called the ``squidadel'' containing the cryogenic RF electronics and quantum amplifier package. This includes quantum noise limited amplifiers, circulators, switches, and temperature sensors. Physical and noise temperatures of the cryogenic electronics housed in the squidadel largely determine the noise temperature of the system, therefore the squidadel is kept thermalized to the dilution refrigerator mixing chamber, the coldest part of the system. 

The small power emitted from the cavity passes through the first series of switches and circulators to the first stage quantum amplifiers and further to the HFET amplifier before passing into the warm space. The MSA boosts the signal with a characteristic gain of $20-25$~dB, followed by an HFET amplifier in the $10$~K space providing a boost of $30$~dB. The squidadel is wired completely with copper coaxial cables whereas wiring from main port antenna to the first stage quantum amplifier is NbTi. Coaxial cables in the input chain are stainless steel. Characterizing the early-stage electronics is an extremely important procedure in determining the sensitivity of the experiment and is covered in Section~\ref{sec:Preparations} as well as in \citep{Khatiwada2020}. The total gain in the cold space is $50-55$~dB.

\subsection{Warm Receiver Chain}

\begin{figure*}
\centering
\includegraphics[width=0.9\textwidth]{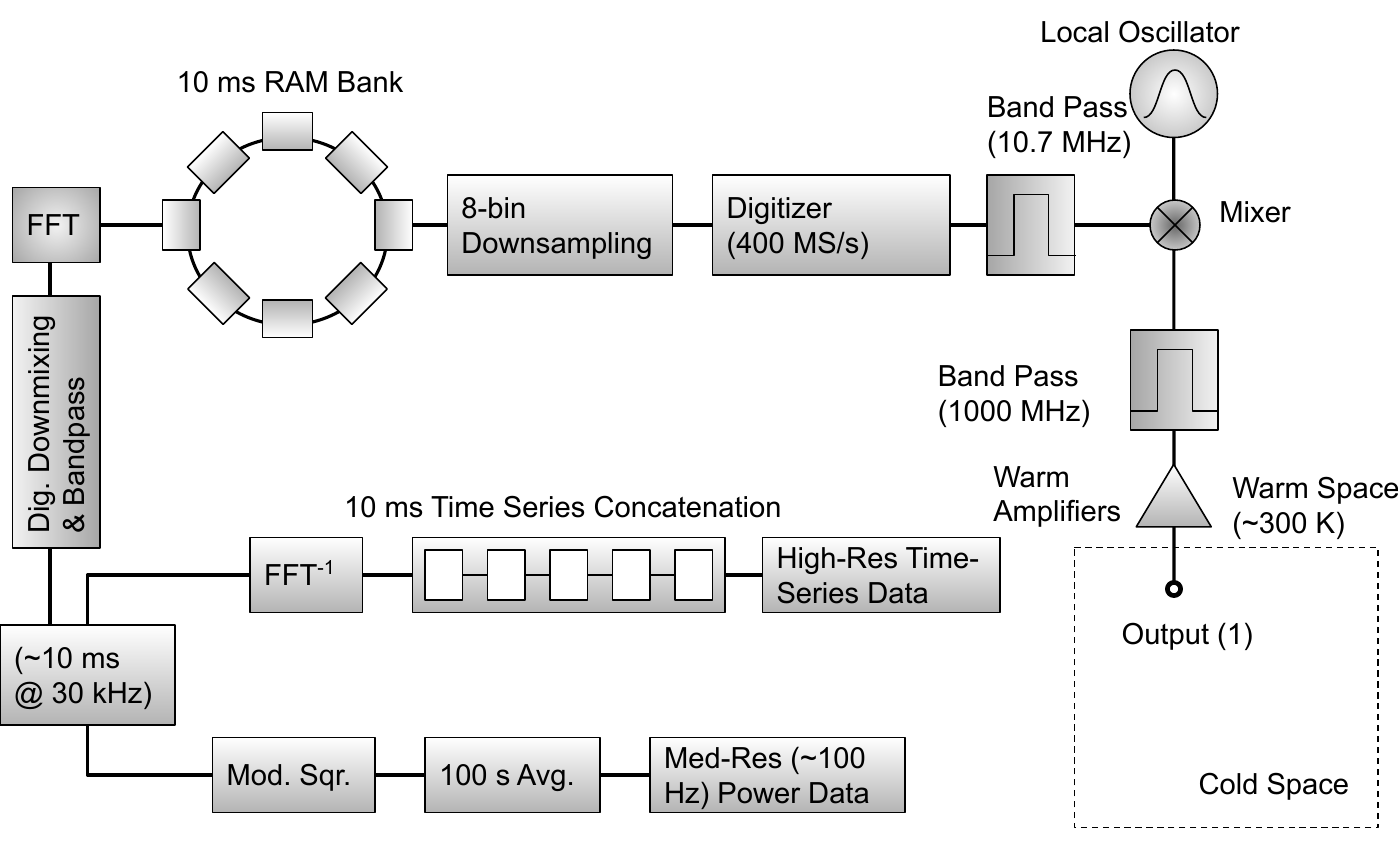}
\caption{Schematic of the Run 1A digitizer chain. Note that Output (1) of the cold RF receiver shown in Fig.~\ref{fig:rec_chain} is used as the input of the warm space amplification and digitization chain. The raw data from a $100$~s integration is recorded in two forms: as a power spectrum with $\approx 100$~Hz wide bins and as a 25M element voltage time series.}
\label{fig:digitizer}
\end{figure*}

The remainder of the receiver chain runs through the successively warmer cryogenic spaces and vacuum feeds into the room temperature space. The total room-temperature amplification is approximately $40$~dB, bringing the overall expected power to the pico-watt level. This power is then directed to a variable-frequency super-heterodyne receiver for digitization, as seen in Fig.~\ref{fig:digitizer}. The power and gain of the receiver chain with the MSA switched out of the system were measured as a function of frequency every few weeks and found to change by a negligible amount, below the 1\% level.

Treatment of the insert RF emissions into a digitized data set is performed in several steps, see Fig.~\ref{fig:digitizer}. The first step is to mix the signal such that the mode center frequency is centered at $10.7$~MHz via a local oscillator set to $f = \nu_0 + 10.7$~MHz.
The down-mixed voltage is bandpass filtered in a $30$~kHz window centered at $10.7$~MHz, which is expected to cover at least the full width at half maximum of the TM$_{010}$-like mode. The analog signal now exists only in this vicinity of $10.7$~MHz and is capable of being sampled quickly enough to resolve structures well below the expected total axion signal width. Time-wise sampling then occurs at a rate of $400$~Mega-samples/s with a 10-bit digitizer \citep{Khatiwada2020}. To optimize the resolution/precision function of potential analyses, digitized samples are partitioned into bins $8$ samples wide and averaged. This down sampling and averaging improves the signal to noise of samples by $\sqrt{8} \sim 3$ times and increasing the bit depth by $log_4(8) = 1.5 $. The resulting re-binned sample has an effective width of $25$~MHz, well above the $2 \times 10.7$~MHz Nyquist rate of the central frequency. The re-binned time series are split into $\approx 10$~ms blocks and temporarily stored to a circular RAM queue. No safety mechanisms exist to preserve the oldest unprocessed sequences, which are overwritten by the newest recordings. If overwritten, those data are lost and a flag is raised in the integration's metadata, indicating that scans processed are not necessarily consecutive. This becomes important for the high-resolution stored data. Once the $10$~-ms series blocks are pulled off the RAM buffer, a Fast Fourier Transform (FFT) is performed on the concatenated series to $12$~MHz resolution, near the Nyquist limit. With the well-resolved contributions about the resonance centered at $10.7$~MHz, a digital mixer centers a $30$~kHz band and shifts these contributions down to begin at $0$~Hz, removing contributions above $30$~kHz with a low-pass filter. We now have drastically down sampled the coherent frequency-space data set of a $10$~ms sample at a rate resolving the full $30$~kHz. The high resolution time series data is formed by performing and inverse FFT on each $10$~ms sample and concatenating the time series in order. The medium resolution (MR) power spectral density, or power spectrum, of the scan is calculated from the modulus squared of each $10$~ms sample, averaged over the 100~second scan ($\sim 10^4$ samples). The power spectrum array then has its edge bins removed, producing the 256 bin wide form of the raw spectrum saved for the MR analysis. The width of each bin the MR spectrum is $ \approx 100$~Hz, the Nyquist limit for each $10$~ms sample. It is in this power spectrum that the axion dark-matter signal would appear as a localized excess.

\subsection{Experiment Status Measurements}
\label{sec:statusreadings}

The warm receiver chain and digitizer are located within the larger structure of the ADMX DAQ system. A number of measurements are recorded by the DAQ to monitor the state of the insert, magnet, and ancillary cryogenics. Interpretation of these experimental conditions plays a critical role in performing the offline analysis. Here we briefly describe the measurement and interpretation of these experimental conditions. A more detailed account can be found in \citet{Khatiwada2020}.

The data recorded from the experiment can be divided into periodically sampled experimental state information and radio-frequency measurements taken in the course of the axion search. The experimental state information consists of status readings from temperature, pressure, field, and current sensors.

For temperatures above $1$~K, an assortment of resistance sensors are used to read out temperatures at various thermal stages. For temperatures below $1$~K, temperature sensors are sampled with a four-wire resistance measurement using a Lakeshore Alternating Current (AC) resistance bridge \citep{Khatiwada2020}.
The temperatures of the cavity and quantum electronics package are measured using Cernox temperature sensors and the temperature of the mixing chamber mounted to the cavity is measured using a Ruthenium Oxide sensor \citep{Khatiwada2020}. During operations, it was observed that the Cernox sensors had a large magneto-resistance at temperatures below $1$~K. With the magnet ramped to $6.8$~T, The Cernox temperature sensors on the main cavity read $70\%$ higher temperatures compared to the magnet ramped down, while the Ruthenium Oxide temperature sensor on the mixing chamber increased by $2\%$. Thus, in Run 1A, the temperature of the cavity was read by the Ruthenium Oxide temperature sensor mounted to the mixing chamber. Because the quantum electronics package was kept in a field-free region, Cernox temperature sensors located on the package did not suffer any appreciable effects from the magnetic field, and were used to measure the physical temperature of the quantum amplifier.

The main magnet state is captured by several sensors on the magnet's power supply as well as Hall probes that directly measure the magnetic field parallel to the probe wire, which are set in an azimuthal (vertical) orientation in areas of high magnetic field as well as in low-field/high-sensitivity areas like the so-called field-free region. The power supply monitors the voltage and amperage being fed to the main magnet at a sampling rate of a few minutes. As stated by the manufacturer, the peak magnetic field for an empty bore can be calculated from Eqn.~\ref{magnetBmax}, though minimal impact from insert materials and bucking coil are expected near the cavity. The Hall probes record the magnetic field with a period of about an hour. The magnetic field throughout the cavity may be modeled using the combined data and then used to compute the form factor.

\section{Data-taking Operations}
\label{sec:Operations}

Data-taking operations for Run 1A occurred between January 18 and June 11, 2017. During that time, the TM$_{010}$-like mode of the main cavity was tuned over the range of $645-680$~MHz and observed for power excesses consistent with an axion DM signal. The identification and handling of candidates was also performed during this period. This section decomposes the data-taking process into its constituent parts and cadences down to the level of a single RF digitization, overviewed in the previous section.

\subsection{Overall Structure of data-taking}

\begin{figure}
\centering
\includegraphics[width=\linewidth]{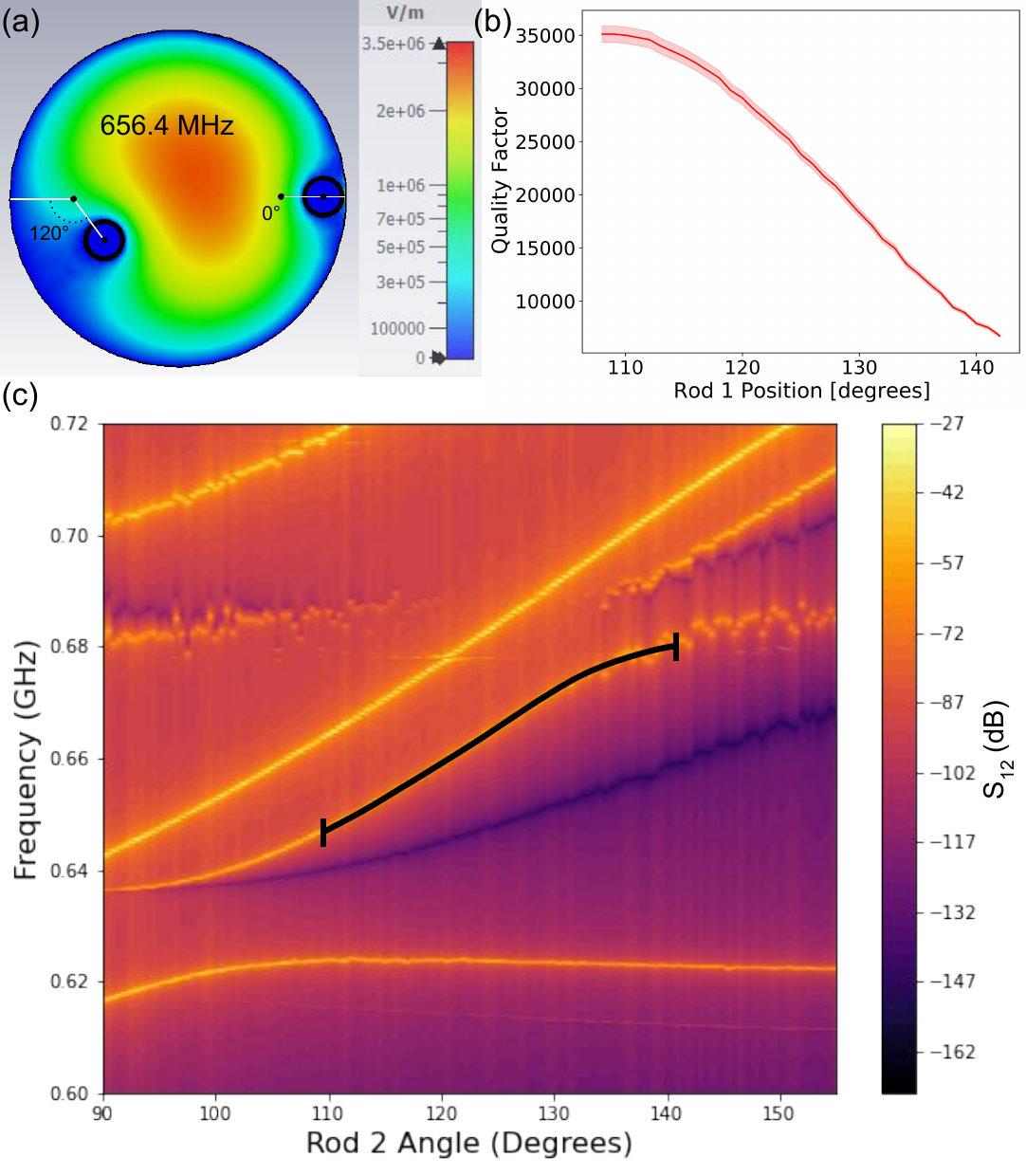}
\caption{Tuning configurations of Run 1A: (a) The electric field strength component $\vec{E} \cdot \hat{z}$ for an example rod configuration for Run 1A included one rod being placed at the wall while the other turned from $107-141$~degrees; (b) measured cyrogenic $Q_L$ of the TM$_{010}$-like over the rod tuning range; (c) simulated transmission (S$_{12}$) spectra of the cavity, highlighting modes between $500-1000$~MHz over the rod tuning range. The target TM$_{010}$-like mode over the Run 1A scanning range is bracketed in black.}
\label{fig:rod_tuning}
\end{figure}

The global imperative for data-taking in Run 1A was to cover the viable frequencies at the intersection of the cavity and receiver chain operational ranges. Once at sub-Kelvin temperatures, the MSA was established to be the limiting device in the receiver chain, providing the target gain over the range $645-680$~MHz, quickly decaying for lower and higher frequencies. This range coincides with a continuous stretch of the TM$_{010}$-like cavity mode with high form factor $C$, high quality factor $Q$, and without significant mode crossings. The frequency range is accessed through tuning rod configurations where one rod is held at the wall and the other is free to turn, as seen in Fig.~\ref{fig:rod_tuning}.

The $35$~MHz range is divided into four segments, observed chronologically as: $650-660$~MHz; $660-670$~MHz; $670-680$~MHz; and $645-650$~MHz. Each segment is individually scanned to DFSZ sensitivity, investigated for candidates, and cleared of candidates before moving onto the next segment. The range is modularized for faster turn-around in the case of a prominent candidate, and for feedback on the state of the MSA.

The scan process of each segment is structured in a sequence of up-tuning and down-tuning sweeps of the TM$_{010}$-like mode over several weeks to provide uniform coverage. Multiple sweeps separated in time by days or weeks also provides an opportunity to sample possible periodic and transient behavior of the axion field. This cadence is only rarely broken in the case of a promising candidate. 

Once the scans were complete to DFSZ sensitivity according to the live analysis, the procedure transitioned into candidate handling. DFSZ sensitivity is established based on criteria covered in sections Sec.~\ref{sec:Statistics}. The investigation of candidates and the handling decision tree was established in the Candidate Handling Protocol covered in Sec.~\ref{sec:Candidates}. Once candidates for one segment are handled, the process begins again for the next segment.

\subsection{Cadence of a data-taking Cycle}

The process of scanning a segment is broken up into a smaller cadence called a data-taking cycle. The data-taking cycle lasts for approximately $150$~seconds and includes operations of moving tuning rods, active cavity measurements, and the passive integration of RF chain emissions. The data-taking process at and below this level is automated and was often left in continuous and unhindered operation for days barring a necessary manual bias of the MSA.

At the head of the data-taking cycle are several active measurements of the cavity and receiver chain using the VNA. The injected sweep signals are far more powerful than a potential axion signal or impinging RFI, making them easier to see, while still conforming to the receiver operating parameters optimized for the passive integrations. The first measurement is a swept transmission S$_{12}$ from the weak port through the receiver including the cold and warm amplifiers. This measurement shows the cavity response, yielding a measurement of mode frequencies and quality factors. S$_{12}$ measurements take approximately $10$~seconds each. A wide-band transmission scan is made every 10-th scan cycle for mode mapping purposes. The transmission measurement is followed by a reflection measurement taking approximately $20$~seconds, where the signal is sent through the bypass line and is directed towards the cavity by a circulator (C$_1$ in Fig.~\ref{fig:rec_chain}). The reflection sweep near the resonance is mostly absorbed by the cavity, while off resonance the signal is reflected and passes back through the cold and warm amplification of the RF system. This measurement yields a wide-band measurement of system gain for noise calibrations, the coupling between the antenna, and the cavity mode of interest, as well as the $Q_L$ of the cavity mode of interest. Coupling of $\beta=1$ is consistent with critical, or impedance-matched, coupling. The duration of transmission and reflection measurements have been reduced to less than a second each in subsequent runs by optimizing input attenuation and power \citep{Bartram2021}.

The receiver integrates emissions from the cold RF for the remainder of the data-taking cycle, which comes to an observation duty factor of approximately $0.7$.  
At the end of the data-taking cycle, there are two mechanical tuning processes to prepare for the next integration. The tuning rate of the rod can be estimated given a target sensitivity and regional values for the system temperature, quality factor, etc., or can be set manually to a given number of steps per cycle. The tuning rate of the warm stepper motor was set between $0.1-0.2$~radians per cycle during the first passes of a section's bandwidth, which translates to $1-2$~kHz per cycle. The main port antenna is also given the opportunity to tune to alter the coupling to the main cavity. Note that rod and antenna tuning must occur slowly to avoid overtaxing the dilution refrigerator.

\subsection{Logging of Search Data}

Each data-taking cycle is stored as an independent entry in an SQL database on the DAQ's main control computer. Each cycle contains its own unique serial number and timestamp when the cycle is initiated. The integrated power spectrum and a collection of markers are added to the cycle entry to more easily identify the conditions surrounding the experiment. Each entry contains the sum total information necessary to perform an axion search.

\section{Data Preparation for Axion Search}
\label{sec:Preparations}

With data collected, preparations are then made to characterize the state of the experimental apparatus and assess the quality of the measurements to search for the axion. This includes the verification of the cavity magnetic field, characterization of the receiver chain as measured by its effective noise and gain properties and its persistent background structure. This section details how the measurements necessary to axion search are interpreted from their recorded state into actionable data.

\subsection{Direct Measurements}

Temperatures within the experiment are measured by arrays of sensors placed at every level of the insert and on the main magnet casing, both in areas of high magnetic field and in the field-free region. Voltage time series from the sensor leads are interpreted through EPICS and converted into temperature readouts. Sensors of differing makes and models were tested against one another in and out of the magnetic field in order to study the uncertainties and biases present during data-taking operations, as detailed in \citep{Khatiwada2020}. The errors in temperature sensors are reflected in Table~\ref{table:errors}, and are integrated in the final sensitivities and limits.

The magnetic field in the main cavity is computed from the current supplied to the main magnet and confirmed by the Hall probes placed throughout the insert. The maximum magnetic field in the solenoid center is computed from the current via Eqn.~\ref{magnetBmax} and modeled in form by numerical simulation as indicated in \cite{Khatiwada2020}, including the counter field induced by the bucking coil surrounding the field-free region.

Active transmission and reflection measurements taken during each data-taking cycle are analysed to assess the impedance match between the cavity and main port, to match the transmission function through the port to a Lorentzian distribution, and to extract the quality factor and mode central frequency. During Run 1A, the weak port was partially dislodged and became decoupled from the cavity during the insertion and commissioning process, marginalizing the effectiveness of the cavity transmission measurements. As a result, S$_{11}$ reflection measurements were used to assess coupling $\beta$ and loaded quality factor $Q_L$ of the TM$_{010}$-like cavity mode. Recall that the normalized reflection power spectrum is modeled by the form given in Eqn.~\ref{eqn:cavity_reflection}.
The logged S$_{11}$ measurement is fit to this form by a least-squares analysis of the swept spectrum and the parameter values are used to assess the loaded $Q_L$ and coupling $\beta$ of the cavity-main-port state, see Fig.~\ref{fig:mode_shape}.

\subsection{Axion Power in Cavity and Main Port Transmission}

The figures central to the deposition  of axion power into the cavity and transmission into the receiver can now be computed. The power-per-axion density deposited in the cavity at a single frequency is given by
\begin{equation}
   \frac{ \left< P_a \right> }{\left< \rho_a \right>} \propto C_{010} ~ \nu ~ B_{max}^2 ~ V ~ Q ~ T_{\nu_o} (\nu).
\end{equation}
The form factor of the TM$_{010}$-like mode $C_{010}$ is computed from Eqn.~\ref{eqn:modeformfactor} using numerical models of the magnetic field and mode electric field distributions \cite{Khatiwada2020}. The volume of the cavity is computed from \citep{Yu2004} and is invariant to high precision during low temperature operations. The maximum magnetic field is computed from Eqn.~\ref{magnetBmax}, and has error calculated at the level of the power supply current stability. The numerical error of $B_{max}^2 \times C_{010} \times V$  is available in Table~\ref{table:errors}.

The loaded quality factor $Q_L$ is computed from the reflection measurement detailed in the previous sub-section and uses a rolling average and uncertainty from the previous 10 measurements. The portion of the power transmitted from the cavity into the main port, and past the first low-pass filter is modulated by the coupling strength factor $\beta$, and the low-pass filter transmission function $T_{filter}$
\begin{equation}
    \left< P_{port} \right> = T_{filter} ~ \frac{\beta}{1 + \beta} ~ \left< P_{cav} \right>,
\end{equation}
where the filter is near-transparent at the Run 1A frequencies ($T_{filter} \approx 1$).

\subsection{Receiver Gain and Noise}

The remainder of the receiver chain transmits the cavity output, but also imprints its own structure into the digitized power spectrum down to the scale $ \nu_o/2 Q_L \sim 10$~kHz (see \citep{Daw1998,Khatiwada2020}), an order of magnitude wider than the expected axion signal width of $\lesssim 1$~kHz. Directly modeling the total transmission function prior to Run 1A proved to be too unreliable due to high variability in the response of devices and strong inter-device couplings. This subsection analyzes the receiver transmission heuristically to characterize its gain structure and noise.

The noise background is expected to be overwhelmingly thermal. Noise in the power spectrum is first contributed by the fluctuations about the mean photon occupation function of the cavity
\begin{equation}
    \left< n_{\gamma}(\nu) \right>  = \frac{ 1}{e^{ h \nu / k_{\text{B}} T} - 1}
\end{equation}
where $k_{\text{B}}$ is the Boltzmann constant, $T$ is the physical temperature of the thermal source, $h$ is Plank’s constant, $\nu$ is the excitation frequency. The expected power spectrum per frequency of the cavity is derived from the mean energy spectrum
\begin{equation}
    \left< E(\nu) \right>  = -\frac{d log (Z)}{d (k_{\text{B}} T)} = \frac{h \nu }{2} + \frac{ h \nu}{e^{ h \nu / k_{\text{B}} T} - 1}
\end{equation}
where $Z$ is the canonical partition function. Note that the first term of the mean energy spectrum is the vacuum contribution to the cavity energy and the second term is the contribution from non-trivial occupation. The emission of photons out of the cavity, ignoring reflections and attenuation for now, is then given by the free flow of power out of the strong port. The expected power spectrum per frequency is
\begin{equation}
\left< \frac{d P_n(\nu)}{d \nu} \right> = \left< E(\nu) \right>  - E_{vac}(\nu) = k_{\text{B}} T \times \frac{h \nu / k_{\text{B}} T}{e^{ h \nu / k_{\text{B}} T} - 1} \label{eqn:BBodyPower}
\end{equation}
In the limit of the photon energy much less than the bath temperature $h \nu \ll k_{\text{B}} T$, the expected spectrum density flattens and the power spectrum becomes
\begin{equation}
    \left< P_n(\nu) \right>  = k_{\text{B}} T b, \label{eqn:radiometermean}
\end{equation}
where $b$ is the integrated bandwidth and $\nu$ is taken as the center frequency.

The distribution of power fluctuations can be found by looking at the occupation probability of individual states, given by the density function
\begin{equation}
    \rho_r (\nu) = \frac{e^{-(r+1/2) h \nu / k_{\text{B}} T}}{Z} , \label{eqn:densityfunc}
\end{equation}
where $r$ is the occupation number of the state(s) with frequency $\nu$. The emission rate of photons over the bandwidth $b$ is expected to be $\sim k_{\text{B}} T b/ h \nu$. This comes to $\sim 16$~bandwidth-emissions for a typical Run 1A temperature of $T=500$~mK at a center frequency $\nu=660$~MHz. Each short continuous $\delta t \approx 10$~ms integration (bandwidth $b = 1/t \approx 100$~Hz) would then produce an MR emission power spectrum in an exponential distribution of similar shape to Eqn.~\ref{eqn:densityfunc}. The rate parameter of an exponential distribution is identified with the variance square root $\lambda = \mu = \sigma$ where the mean of the power spectrum is already known to be $\mu = \left< P_n \right>  = k_{\text{B}} T b$ from Eqn.~\ref{eqn:radiometermean}. 

The $\Delta t \approx 100$~second integration period of each data taking cycle is made up of $n \Delta t/\delta t \approx 10^4$ MR power spectra, which are considered identically distributed over that time scale and independent as the integration time exceeds the timescale of power fluctuations $\delta t > 1/2b$. Averaging those spectra will shift the distribution shape from an exponential to a normal shape in the large-$n$ limit according to the central limit theorem, leaving the mean unchanged but re-scaling the standard deviation by a factor of $1/\sqrt{n}$. Therefore, we find the distribution of emitted power fluctuations to have standard deviation
\begin{equation}
    \sigma_{P_n} = k_{\text{B}} T b/\sqrt{n} = k_{\text{B}} T \sqrt{\frac{b}{n \delta t}} = k_{\text{B}} T \sqrt{\frac{b}{\Delta t}},
\end{equation}
which matches the Johnson spectra \citep{Johnson1928,Dicke1946}.

The width of the noise distribution is proportional to the thermal temperature, as is the mean thermal power. One can construct a bin-wise signal-to-noise ratio as a proxy to the sensitivity of the instrument to detect a signal of known power
\begin{equation}
    SNR = \frac{P_{\text{signal}}}{k_{\text{B}} T_{sys} b} \sqrt{\frac{\Delta t}{b}} = \frac{P_{axion}}{\left< P_n \right>} \sqrt{\frac{\Delta t}{b}}
\end{equation}

The distribution of power fluctuations through the remainder of the receiver chain is expected to remain thermal, therefore a system noise temperature is used to characterize the net distribution of noise in the receiver. The details of computing the system temperature can be found in \citep{Khatiwada2020}, but we explain them briefly here as they will enter into the gain structure characterization in the next sub-section.

Amplifiers and attenuators in the receiver chain modify their inputs, both signals and noise powers, as well as inject noise power according to their own blackbody spectrum. A simple composition model for the power emitted at the end of the chain is
\begin{equation}
    P_{sys} = \left( ... \left(P_{cav} G_1 + P_{n_1} \right) G_M +  ... + P_{n_M} \right) 
\end{equation}
where $P_{n_i}$ and $G_i$ respectively are the noise power and the gain of the i-th component of the receiver. Assuming thermal power spectra at each stage ($\left< P_{n_i} \right> = k_B T_{n_i} b$), the total system noise can be computed as a temperature by dividing out the total gain of the receiver $G_T = G_1 G_2 ... G_M$
\begin{equation}
    T_{sys} = T + T_{1} + \frac{T_{2}}{G_{1}} + \frac{T_{3}}{G_{1}G_{2}} + ... \label{eqn:Tsys}
\end{equation}
If the gain of the first components are large, then one can effectively truncate the series after the first several terms. Note that the contributions of the power spectrum fluctuations also follow this relation
\begin{equation}
    \sigma_{P_{sys}} = \sigma_{P_{cav}} + \sigma_{P_{1}} + \frac{\sigma_{P_{2}}}{G_{1}} + \frac{\sigma_{P_{sys}}}{G_{1}G_{2}} + ...
\end{equation}

The gain of the MSA varied with both its center frequency and physical temperature throughout the main data run, as did the HFET amplifier to a lesser degree, requiring periodic calibration to optimize the system temperature. Contributions from amplifiers and attenuators downstream of the HFET were measured to be effectively constant. Also during the run, the switch for the heated load malfunctioned, so the primary noise calibration came from an ``on-off resonance'' method, described below and in \citet{Du2018,Khatiwada2020}.

The physical temperature of the cavity in Run 1A ($\sim150$~mK) was significantly different than the physical temperature of the milli-Kelvin electronics ($\sim300$~mK). This difference implied that the relative thermal power on and off resonance encoded sufficient information to determine the system noise in the same way a heated load measurement does. The method for calibrating $T_{sys}$ during Run 1A then became:
\begin{enumerate}
\item Use a ``hot-cold'' measurement as the primary calibration to determine $T_{sys}$ absolutely at a few fixed frequencies, as well as the noise contributions from the RF components downstream of the MSA
\item Use a faster ``SNR-Improvement'' measurement to determine $T_{sys}$ on every digitization
\end{enumerate}
The ideal hot-cold method measures the power at the receiver in a bandwidth of multiple $Q$-widths about the resonant frequency of the cavity while the antenna is critically coupled. The powers measured on and off resonance are then compared to determine the noise contributions of the MSA amplifier from:
\begin{equation*}
R = \frac{T_{attenuator} + T_{amps}}{T_{cavity} + T_{amps}}
\end{equation*} 
where $T_{attenuator}$ is the physical temperature of attenuator A as seen in Fig.~\ref{fig:rec_chain}, $T_{cavity}$ is the physical temperature of the cavity, $T_{amps}$ is the noise contribution from all the electronics of the receiver chain, and $R$ is the ratio of power off-resonance to the power on-resonance. Temperatures for the cavity and attenuator were taken from recordings of the Ruthenium Oxide temperature sensors closest to it \cite{Khatiwada2020}. The $R$ factor can be determined by a model of the cold RF system and fitting a model to the power spectrum as a function of frequency, discussed in \cite{Du2018,Khatiwada2020}.

\begin{figure}
\centering
\includegraphics[width=\linewidth]{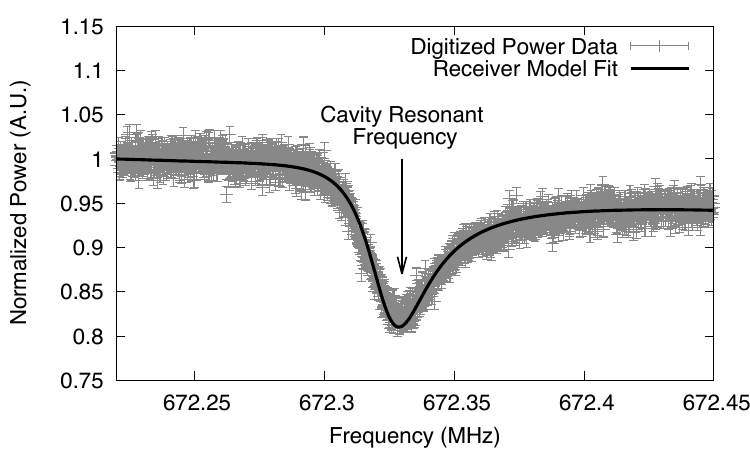}
\caption{Example power measurements used to calibrate the system noise temperature Noise power on versus off resonance acts as an effective hot-cold load. The asymmetry of the shape is a result of interactions between RF components. This figure first appeared in \cite{Du2018}.}
\label{fig:OnOffNoiseMeasurement}
\end{figure}

Parameters had to be changed slightly to accommodate the typical conditions for this  \textit{in-situ} measurement for Run 1A. Off-resonance noise power is dominated by the attentuator (physical temperature typically $\sim 300$~mK) and also contains a contribution from the receiver noise temperature. On-resonance noise power is the sum of the cavity (physical temperature typically $\sim 150$~mK) and the receiver. The transition between the two creates a dip of order $20\%$ in power seen at the cavity resonance in Fig.~\ref{fig:OnOffNoiseMeasurement}. This measurement yielded a system noise temperature typically of $\sim 500$~mK, but was found to vary substantially throughout the run due to different gains of the MSA, which required frequent re-optimization as discussed in \citet{Khatiwada2020}.

\subsection{Receiver Shape Removal}

As was seen in the previous sub-section, the spectra from a well-equilibrated receiver chain sans persistent signal is expected to be flat, characterized by a single system temperature. This was rarely the case in practice for Run 1A. There are multiple causes of a non-flat power spectrum by experimental hardware, including frequency dependant gain variations before and after mixing down the target frequency, and frequency-dependant noise variations. The last of these is expected to be suppressed as the early receiver chain is nearly homogeneous and stable in temperature over a data taking cycle. 

It is crucial to know the structure of the gain so that it is not confused with potential axion signals. Removing the structure and flattening the receiver power spectrum allows for a more straightforward interpretation of the data as Johnson spectra dominant. The flattened spectra also has computational advantages for the analysis as the gain response of an incoming axion signal is similarly flattened, producing a convolutional filter form to the optimal signal search, a topic that will be covered in Section~\ref{sec:Statistics}. The operational goal of receiver shape removal is to completely remove the frequency dependant receiver chain response while retaining potential signals and thermal components.

Finding the true gain response of this complex system is a difficult task to perform from status measurements of components alone, and has met with limited success \citep{Daw1998}. Techniques rooted in purely heuristic fitting of the power spectra have had far greater success \citep{Brubaker:2017rna,Asztalos2010,Daw1998}. Run 1A used filters of Savitzky-Golay (SG) type and a low order polynomial fits for the off-line and live analyses respectively. 

\subsubsection{Polynomial Filter}

A sixth-order polynomial is used by a live analysis process run in parallel with data-taking operations to fit the receiver structure for computational speed. The live analysis is responsible for providing a real-time feed of the experiment's sensitivity and for identifying potential candidates. 

\begin{figure*}
\centering
\includegraphics[width=\linewidth]{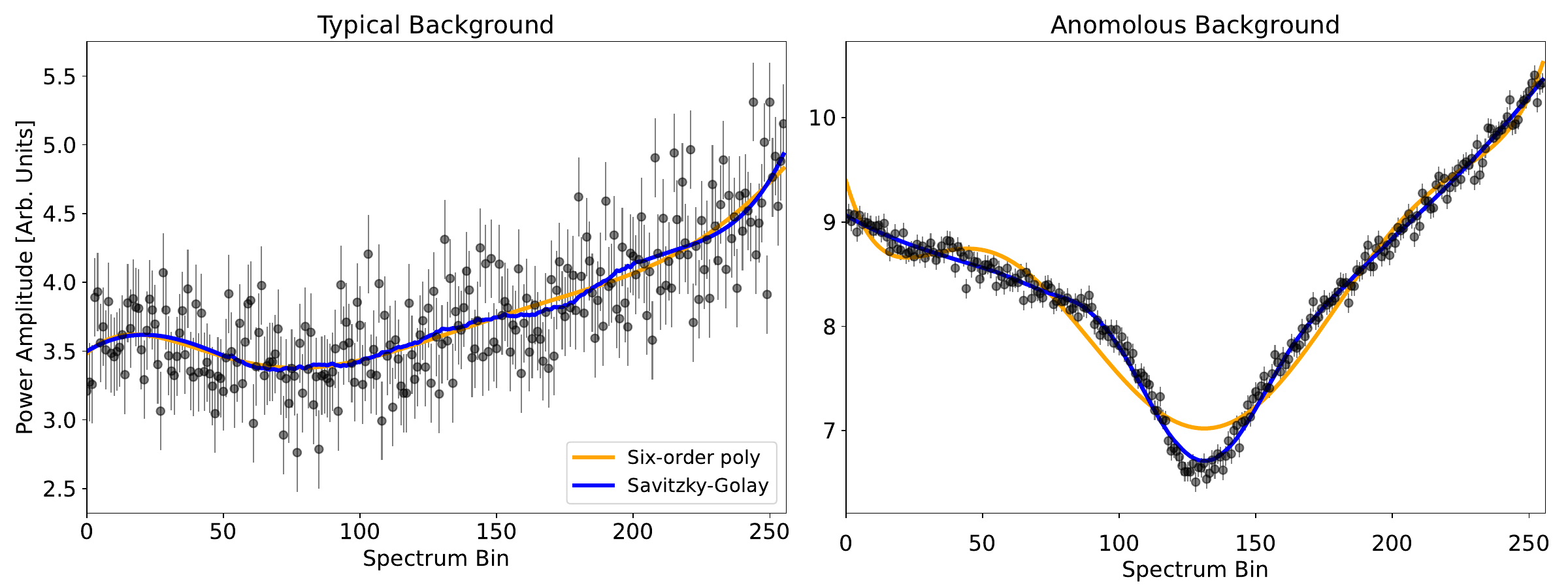}
\caption{Sample fits to two observed receiver spectra by a six-order polynomial and Savitzky-Golay windowed spline. The two raw spectra represent the typical (left) and anomalous (right) imprints of receiver structure. The polynomial is adequate to fit small and low harmonic structure to the noise floor, but fails to respond to large structure. Savitzky-Golay is seen to provide a more consistent fit over the range of experienced structures in Run 1A, though artifacts of the spline fit can be seen in the form of fluctuations for scales at and below the window size. Such fluctuations lead to small negative correlations in the background-subtracted spectra.}
\label{fig:backgroundfit}
\end{figure*}

The polynomial fit works well for spectra that are already nearly flat, but its quality degrades significantly over spectra containing larger and more complicated variations, as seen in Fig.~\ref{fig:backgroundfit}. There were multiple periods during Run 1A, either due to the operating state of the MSA or other elements, where the receiver structure was highly perturbed. A low-order polynomial fit does a poor job of conforming to large scale fluctuations at multiple harmonics \cite{Boutan:2017oxg}. This shortcoming resulted in some background-subtracted scans with residual structure above the noise floor, resulting in an excess of axion candidates that impacted rescan operations. The candidate procedure will be discussed more in Sec.~\ref{sec:Candidates}.

\subsubsection{Savitzky-Golay Filter}

The SG filter was used to fit the receiver structure after data-taking operations were complete to conduct the analysis that provided the limits in \citet{Du2018}. The SG filter proved to be much more versatile over the range of structures experienced in Run 1A, as seen in Fig.~\ref{fig:backgroundfit}. The parameters used for the SG filter were $d = 4$ for the spline order and $L = 121$ for the size of the box window function. These values were chosen as they produced adequate fits to a wide range of observed backgrounds with minimal impact to potential axion signal shapes.

The general thinking is that you want to choose filter parameters that fit typical background shapes well, but are unable to fit signal shapes well.  I'm not sure I can give why precisely those numbers were used, but my recollection is that they were tuned to minimize our 'fudge factor', the inefficiency (as measured on synthetic signals) introduced by background subtraction.

\begin{figure*}
\centering
\includegraphics[width=\linewidth]{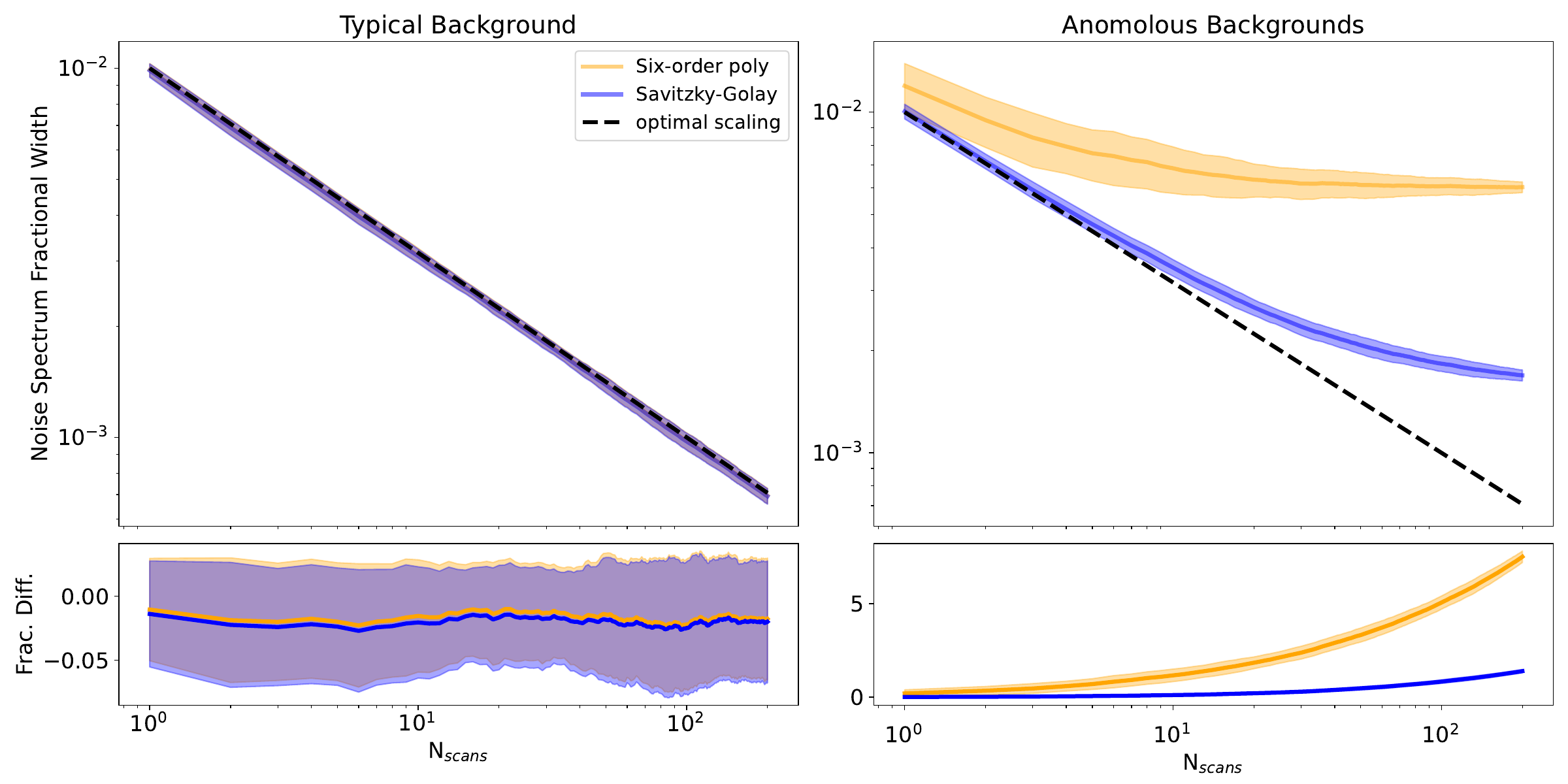}
\caption{Scaling of fractional fluctuation distribution width with number of co-added signals for randomly generated simulated receiver spectra with typical shape (left panel) and anomalous receiver shapes (right panels). Solid lines indicate the average width, while the shaded band gives the one sigma uncertainty. The ideal variance scaling of co-added uncorrelated noise is $\sigma^2/\mu^2 \propto 1/N_{scans}$ (top panels dashed line). Fractional differences (lower panels) are relative to the ideal scaling.}
\label{fig:coadditionscaling}
\end{figure*}

The SG filter has been found to attenuate axion signals and imprint small negative correlations between processed spectrum bins \cite{Brubaker:2017rna}. These anti-correlations become problematic for scans repeated over the same frequency range under the same conditions, such as may occur for a re-scan, as seen in Fig.~\ref{fig:coadditionscaling}. Accumulation of anti-correlations will occur very rarely under normal scanning operations. No new candidates were produced under the SG filter.

\subsubsection{Preparation of a Spectrum}
 
\begin{figure}
\centering
\includegraphics[width=\linewidth]{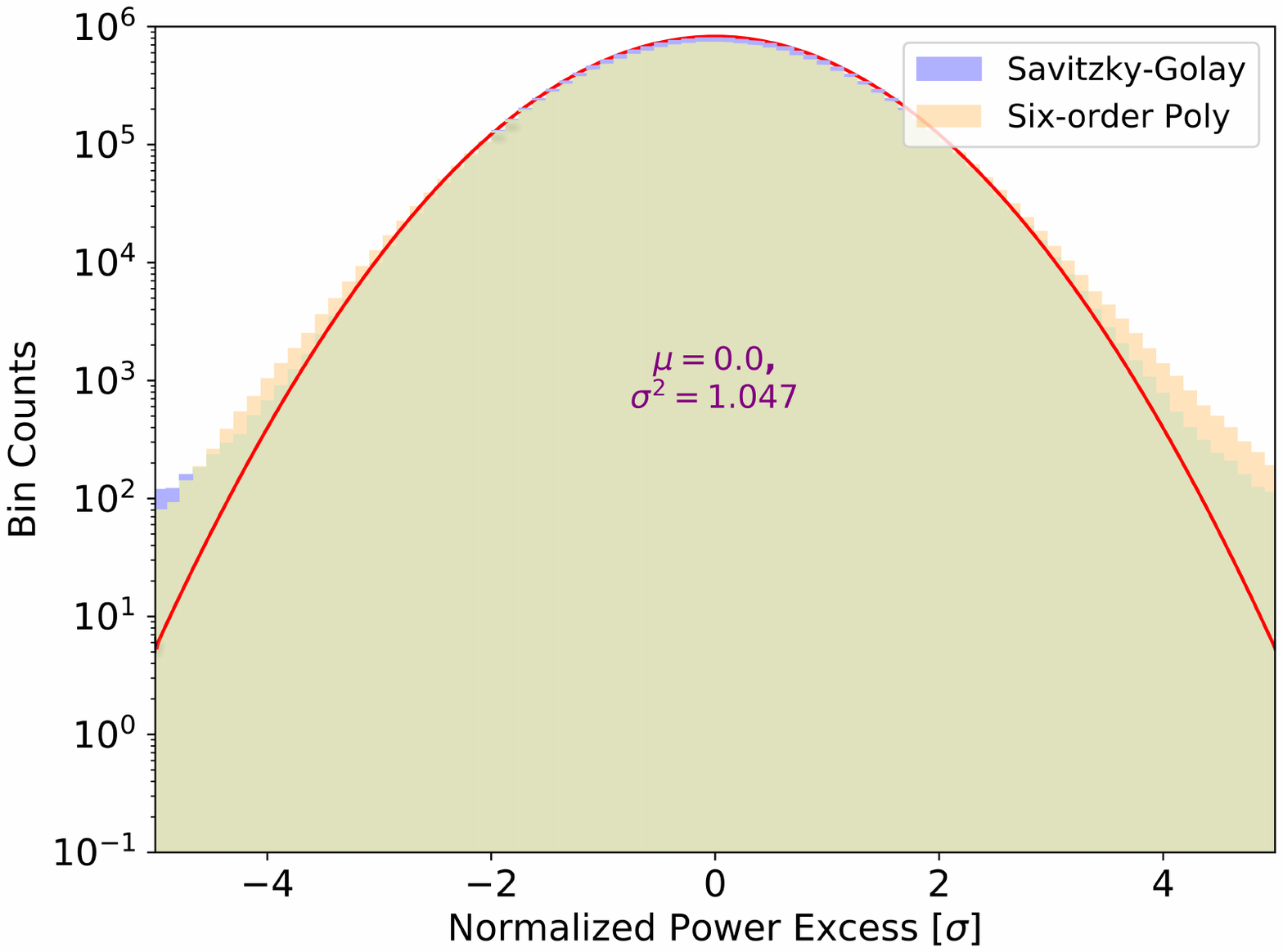}
\caption{Histograms of normalized power fluctuations for all scans used in Run 1A MR analysis after background subtraction using six-order polynomial (orange) and Savitzky-Golay (blue) filters, with green indicating overlap. The fluctuations $\delta$ are normalized to the thermal spectrum width. The Gaussian curve (red line) fitted to expected distribution of fluctuations shows the distribution is well fit at the center but is slightly deformed from a purely thermal spectrum, containing notable excess counts by $\pm 3 \sigma$ for the six-order polynomial background subtraction. The Savitzky-Golay background subtraction technique fares somewhat better, with excesses appearing closer to $\pm 4 \sigma$ from center. Injected power in the form of RFI is responsible for some portion of the positive power excesses, while much of the remaining large power excesses can be attributed to sub-optimal background fitting.
}
\label{fig:noisedistr}
\end{figure}

We may now prepare the fitted scan for axion search. Using the findings of the previous sub-section, we can see that dividing the power spectrum by the background shape produces a dimensionless mean-normalized spectrum 
\begin{equation}
    O_p = \frac{P_{spec}}{P_{fit}}.
\end{equation}
with unit mean and, in the absence of residual background structure, Johnson-distributed fluctuations with uniform variance over the entire frequency band. We are only concerned with power excesses above the mean as our axion signals are of narrow bandwidth, making the relevant spectra 
\begin{equation} 
\delta = O_p - 1 
\end{equation}
The dimensionless fluctuations are of random normal distribution with fractional width relative to the mean calculated to be
\begin{equation}
    \sigma_{\delta} = \frac{\sigma_{P_n}}{\left<P_n \right>} = \frac{1}{\sqrt{b \Delta t}}, \label{eqn:deltaswidth}
\end{equation}
which is expected to apply across the spectra as long as the receiver components up to the first stage amplifier are in thermal equilibrium and the gain structure of the receiver thereafter is much wider than the scan bandwidth. This means one can compare the fluctuation statistics across an MR scan, and between MR scans so long as the scan times match. The dimensionless fluctuations show overwhelming normal distribution shape, as seen in Fig.~\ref{fig:noisedistr}.

To find the unit-full power excesses, recall that the mean of $O_P$ is identified with the average thermal power out
\begin{equation} 
\bar{P} = k_{\text{B}} T_{sys} b.
\end{equation}
Multiplying the fluctuations by the mean power produces the normalized fluctuation power \cite{Du2018}
\begin{equation} 
\delta_w = \delta \times \bar{P}  .
\end{equation}
These are the power fluctuations against which the search for axion emissions from the cavity is performed.

The SG filter was able to fit the background well enough so as to not change the noise width more than a factor of two from the theoretical expectation of Eqn.~\ref{eqn:deltaswidth} for 96.6\% of usable Run 1A scans, compared to 90.6\% for the polynomial filter, providing a more complete sampling of the noise distribution, which was found to be overwhelmingly Gaussian normal in shape, as seen in Fig.~\ref{fig:noisedistr}.

\subsection{Errors and Uncertainties}
\label{sec:Errors}

This sub-section provides the systematic uncertainties for parameters used in the analysis, summarized in Table~\ref{table:errors}. First, the uncertainty on the quality factor was quantified by repeatedly measuring the quality factor in a narrow range of frequencies, in this case, from $645-647$~MHz, where the quality factor was not expected to change much according to models.  The fractional uncertainty on the quality factor in this range was determined to be $2.2\%$. The fractional uncertainty on the main port coupling was also computed over the same frequency range, and determined to translate to a transmitted power uncertainty of $0.55\%$. The  uncertainty of the individual temperature sensors above $1$~K were taken from their data sheets and the sensors below the $1$~K stage were computed as indicated in Sec.~\ref{sec:statusreadings} and \cite{Khatiwada2020}. The combined error on the system temperature $T_{sys}$ is calculated at $7.1\%$ by adding in quadrature the individual error contributions from the components in Eqn.~\ref{eqn:Tsys}, being dominated by the cavity (Scientific Instruments RO600 Ruthenium-Oxide sensor on the mixing chamber) and first stage amplifier (Lakeshore CX-1010 Cernox sensor on the squidadel). The uncertainty of the receiver noise temperature, and therefore the system noise, was primarily given by two contributions: the uncertainty in fit for the HFET noise, and the uncertainty in fit of the MSA noise. The total fractional uncertainty for the system noise amounted to $7.5\%$. Further details can be found in \cite{Khatiwada2020} and the supplementary material of \cite{Du2018}. The final systematic uncertainty of $\pm 13\%$, as shown in Table~\ref{table:errors}, was computed simply by adding all listed uncertainties in quadrature as the errors are assumed to be independent from one another.

\begin{table}[h!]
\centering
\begin{tabular}{ |p{5cm}||p{1.25cm}|  }
 \hline
 \multicolumn{2}{|c|}{Systematic Errors and Uncertainties List} \\
 \hline
 Element & Fraction \\
 \hline
 $B_{max}^2 \times V \times C$ & 0.06 \\
 $Q$ & 0.022 \\
 Coupling & 0.0055 \\
 RF model fit & 0.046 \\
 Temperature Sensor Calibration & 0.071 \\
 System Noise Calibration & 0.075 \\
 Background Subtraction & 0.046 \\
 \hline
 Total & 0.13\\
 \hline
\end{tabular}
\caption{Dominant sources of systematic uncertainty. The uncertainties were added in quadrature to attain the uncertainty on the total axion power from the cavity, shown in the bottom row. For the first entry, $B$ is the magnetic field, $V$ is the cavity volume, and $C$ is the form factor. The RF model fit and temperature sensor calibration discussed in \citep{Du2018}. The total uncertainty from these independent sources was computed from the quadrature sum and is incorporated into the final limit computations in Section~\ref{sec:Limits}.}
\label{table:errors}
\end{table}

\subsection{Data Cuts}

Not all data taken during Run 1A operations is qualified to enter into the axion search. Scans taken as part of
\begin{itemize}
    \item receiver studies,
    \item SAG tests,
    \item un-subtract-able RFI,
    \item mode navigation errors,
    \item abnormal cryogenics conditions,
    \item and incomplete logs
\end{itemize}
are flagged for omission. These flags are either explicitly written to the scan or logged by hand in an electronic log that is updated throughout experimental operations, upgrades, and day-to-day upkeep. Further cuts to the data were made during both the live and off-line analyses due to derived knowledge of the experiment being in a poor, or poorly understood, state. This subsection presents the conditions used to cut data from the axion search analysis and its impact on the number of viable scans.

The Run 1A data set consisted of a total of 173,048 scans and raw integrated power spectra: 138,680 for preliminary scans and 34,396 for leveling and rescans. After implementing the analysis cuts described above and itemized in Table~\ref{table:badscans}, 78,958 scans remained for the axion search analysis.
Motivation for these cuts were as follows. First, quality factors lower than $10,000$ or greater than $70,000$ were omitted from the analysis as they were determined to be either compromised by a mode crossing or non-physical and the result of a poor fit to the reflection scan. System noise temperatures below $0.2$~K and above $5.0$~K were excluded as these were likely a poor fit by the SNRI or other temperature fitting mechanism described above. Temperatures below the lowest physical temperature in the experiment of $0.15$~K are flagged as non-physical. Additionally, the SG fit to the power spectra background in the offline analysis was required to have a fractional standard deviation relative to the radiometer equation between $0.5-1.2$ among $95\%$ of the least deviant points in the spectra. This proved sufficient to reject poor fits while retaining potential axion signals.

\begin{table}[h!]
\centering
\begin{tabular}{ |p{3cm}||p{3.1cm}|p{2.2cm}|  }
 \hline
 \multicolumn{3}{|c|}{Data Cuts Conditions List (In Order)} \\
 \hline
 Parameter & Cut Condition & Scans Removed \\
 \hline
 Eng. \& Rec. Studies & various & 84100 \\
 Mode Freq. (MHz) & $640 \ge \nu \ge 685$ & 422 \\
 Quality factor & $10,000>Q>70,000$ & 289 \\
 System Noise (K) & $0.2>T_{sys}>5.0$ & 6028 \\
 Std. Dev. Increase & $0.5>\sigma>1.2$ & 2816  \\
 Other & various & 435 \\
 \hline
 Total & & 94090 \\
 \hline
\end{tabular}
\caption{Conditions and outcomes for data cuts in the offline analysis. Axion search data collected in these conditions were considered irrelevant or of poor quality. Conditions were applied in top-to-bottom order, making a condition's number of scans removed dependant on the conditions listed above it.}
\label{table:badscans}
\end{table}

\section{Power Excess Search}
\label{sec:Statistics}

This section presents the statistical method used to search for persistent signals such as the axion in the MR integrated power spectra. Likelihood functions for power excesses in a scan, the axion signal, and conditions of the apparatus are first established, followed by searches on single data-taking cycles, which are then combined into a ``grand search spectrum''.

\subsection{Statistical Modeling}

It was established in the previous section that an integrated spectra's background noise is dominated by a thermal spectrum from the receiver chain where the receiver is expected to be in local equilibrium over the course of a single data-taking cycle. Under the integration and sampling rates set in the MR data as discussed in Section~\ref{sec:Preparations}, the $\sim 10^4$-sampled Poisson-distributed thermal background power spectra when co-added ideally form a raw power spectrum with random normal distribution of mean power set by the product of the system temperature and the receiver's total gain $T_{sys} ~ G_{tot,\nu}$, and distribution width of $\sigma  = \mu / \sqrt{b \Delta t}$. Once gain structures and mean power have been removed, the remainder of the spectrum optimally consists only of the random-normal thermal fluctuations and externally induced power excesses.

The signal from the local axion field, broadly speaking, would present itself as a power excess of magnitude set in Eqn.~\ref{eqn:axionpwr} and an expected Q-width $Q_a \sim 10^6$ set by the virial velocity of the Milky Way. Note that the axion signal is present as a power excess as opposed to a deficiency as the conversion of a thermal photon to an axion via an inverse-Primakoff or other process is suppressed by an additional factor of the ratio between local photon and axion occupations $n_{\gamma}/n_a$. This implies that the statistical search for an axion signal can be straightforwardly conducted as a one-sided p-value test with underlying random normal uncertainties.

The likelihood function for an axion model hypothesis test is then taken to be
\begin{equation}
    \mathscr{L} \left( H_{a} | D_{s} \right) \propto \prod_{i=1}^{N} \frac{\exp \left\{ -S_{(a^i|s^i)} / \lambda_{d^i}  \right\}}{\sqrt{\pi \lambda_{d^i}}} \label{totallikelihood}
\end{equation}
where $H_a$ is the axion model hypothesis, $D_s$ is the data set, $S$ is the action of the data and model hypothesis, $\lambda$ is the data uncertainty measure, $i$ is the index over observations, and $N$ is the number of observations. It has already been shown that the integrated power spectra observations are nearly bin-wise independent in their noise background. The action of the data and model hypothesis will be taken over the power outputs from the cavity
\begin{equation}
    S_{(a^i|s^i)} = \left( P_{a}^i - P_{s}^i \right)^2
\end{equation}
where $P_{a}$ is the expected transmitted axion power given by Eqn.~\ref{eqn:axionpwr} and $P_{s}$ is the recorded power spectrum. The expected axion power density $P_{a}$ is dependent on the transmission function of the cavity mode $T_{\nu_o}$, the cavity coupling $\beta$ and quality factor $Q_L$, magnetic field $B$, the observed frequency $\nu$, the local axion density $\rho_a$, and the overall shape of the axion frequency distribution function above the rest mass $f(\nu - \nu_a)$. Note that the variance about the axion power is essentially vanishing in comparison to the expected power spectrum due to the high occupation values in the relic axion condensate, a topic that is discussed in Sections~\ref{sec:Apparatus}, \ref{sec:Signal}. The action is modulated by an uncertainty measure
\begin{equation}
    \lambda_{d^i} = 2 \left( \sigma_{P_i} \right)^2
\end{equation}
where $\sigma_P$ is the uncertainty in the recorded power spectrum and is constant over the scan. The uncertainty in a scan's power,
\begin{equation}
    \sigma_P =\frac{k_{\text{B}} T_{sys} b}{ \sqrt{b \Delta t}}, 
\end{equation}
is given in terms of the expected noise power uncertainty as computed in Eqn.~\ref{eqn:deltaswidth}.
The action and uncertainty measures form a $\chi^2$-like kernel for the likelihood, which takes the form
\begin{equation}
    \chi^2 = \sum_{i}^N \frac{(\delta_{w, i} -  T_{\nu_0, i}\braket{P}_{tot} p_{a,i})^2}{2\sigma_{P_{i}}^2},
\end{equation}
where $i$ is the index over bins in the digitized spectra, $T_{\nu}$ is the cavity mode transmission shape, $\braket{P}_{tot}$ is the total power of the model axion signal, and $p_{a}$ is the probability distribution function of relic axions. To better track the coupling parameter, let us expand the model axion power as a fraction of the DFSZ benchmark model $\braket{P}_{tot} = A \times P_{DFSZ}$, where $A$ is the fraction of the model's power relative to the benchmark and $P_{DFSZ}$ is computed using Eqn.~\ref{eqn:axionpwr} sans the mode transmission shape. Refactoring the $\chi^2$ figure in powers of $A$  
\begin{align}
\chi^2 &= \sum_{i}^N \frac{(\delta_{w, i} - A P_{DFSZ} T_{\nu_0, i}} p_{a,i})^2{2\sigma_{P_{i}}^2} \nonumber \\
 &= \sum_{i} \frac{\delta_{w, i}^2}{2 \sigma^2_{P_{i}}} - 2A P_{DFSZ} \sum_{i} \frac{T_{\nu_0,i} p_{a,i} \delta_{w, i}}{2 \sigma^2_{P_{i}}} \nonumber \\
 &+ A^2 P_{DFSZ}^2 \sum_{i} \frac{T_{\nu_0,i}^2 p_{a,i}^2 }{2 \sigma^2_{P_{i}}}, \label{eqn:chi2}
\end{align}
where the zeroth, first, and second power terms may be referred to as $\chi^2_{A^0}$, $\chi^2_{A^1}$, $\chi^2_{A^2}$ respectively.

There are several priors that need to be declared in order to parameterize the space of incident conditions and hypothesis to be tested. This analysis operates under the following assumptions:

\begin{itemize}
    \item Relic axions dominate the local axion distribution.

    \item Relic axions make up 100\% of the DM. 
    
    \item The likelihood of the axion rest mass $m_a$ is uniform over the Run 1A frequency range.
    
    \item The likelihood of the axion-photon coupling $g_{a \gamma \gamma}$ is uniform over all values covered by this search.
    
    \item Incident conditions of the experiment state are random normally distributed about a data-taking cycle.
    
\end{itemize}

The parameters of interest for the posterior distribution function are the intrinsic axion mass and the axion-photon coupling strength. One can marginalize over all other parameters above to reduce the likelihood function to be over only the parameters of interest. The tests here are organized first over axion masses, then the local distribution of the axion field, and lastly coupling strength. 

Using axion power as proxy for the coupling, the posterior probability function for a given axion mass and local distribution reduces to a Gaussian with mean estimated by
\begin{equation}
    \mu_A = E[A] = \frac{\chi^2_{A^1}}{2 \chi^2_{A^0}} =  \frac{\sum_{i} \frac{T_{\nu_0,i} p_{a, i} \delta_{w, i}}{2 \sigma^2_{\delta{w, i}}} }{P_{DFSZ} \sum_{j} \frac{T_{\nu_0,j}^2 p_{a, j}^2 }{2 \sigma^2_{P_{j}}}},
\end{equation}
and width
\begin{equation}
    \sigma_A = \frac{1}{\sqrt{2 \delta \chi^2}},
\end{equation}
where
\begin{align}
    \delta \chi^2 &= \chi^2(\mu_A+1) - \chi^2(\mu_A) \nonumber \\
    &= (2 \mu_A + 1) P_{DFSZ}^2 \sum_{i} \frac{T_{\nu_0, i}^2 p_{a, i}^2 }{2 \sigma^2_{P_{i}}} \nonumber \\
    &- 2 P_{DFSZ} \sum_{i} \frac{T_{\nu_0, i} p_{a, i} \delta_{w, i}}{2 \sigma^2_{P{i}}}.
\end{align}

The expected value of the axion power given an instance of data is computed from the posterior mean $\mu_A$. The sensitivity of the data to power excursions of the specified shape is set by the posterior width $\sigma_A$.

The expected power's deviation from the null hypothesis of no axion ($P_a = 0$) in units of the distribution width gives a measure of the significance for the potential fit
\begin{equation}
    \Sigma_A = \frac{\mu_A}{\sigma_A} \label{eqn:significance}
\end{equation}
and is the metric used to determine candidacy of axion signals. One may also construct confidence intervals over the posterior distribution from a data instance, though the significance of such statistics will not be made in this section. In Section~\ref{sec:Candidates} we cover the procedure of identifying candidates and delineating their causes from statistical fluctuation, to RFI, to the relic axion field. The creation of limits after candidates are deferred to Section~\ref{sec:Limits}.

The next sub-section computes the above statistics on a single data cycle.

\subsection{Single Scan Analysis}

A single data-taking cycle is the natural choice over which to perform the axion search analysis outlined above. This subsection provides the computational details in searching for the axion in each RF integration. The following subsection presents the means by which the analyses are combined, extending the search over the entire scanned range.

The axion mass parameter is used to serialize the tests on a single scan. Axion masses are only sensible to be tested if there is an overlap between the power spectrum and the assumed axion distribution line shape, so we restrict ourselves to masses less than the upper-most edge of the scan frequency range ($\nu_a < \nu_{\text{max}}$) and greater than the lower-most edge of the scan frequency range minus the width of the 99~\% axion line shape ($\nu_a > \nu_{\text{min}} - \Delta \nu_{99 \%}$). The mass tests may be formed anywhere along this continuous segment of parameter space, though the utility of sampling below the MR bandwidth is marginal. The sampling of masses in the search for candidates is performed on the resolution of the MR bins, $\sim 100$~Hz, with a starting point of $645$~MHz. The final limits in Section~\ref{sec:Limits} will be averaged over many adjacent mass samples. 

\begin{figure}
\centering
\includegraphics[width=\linewidth]{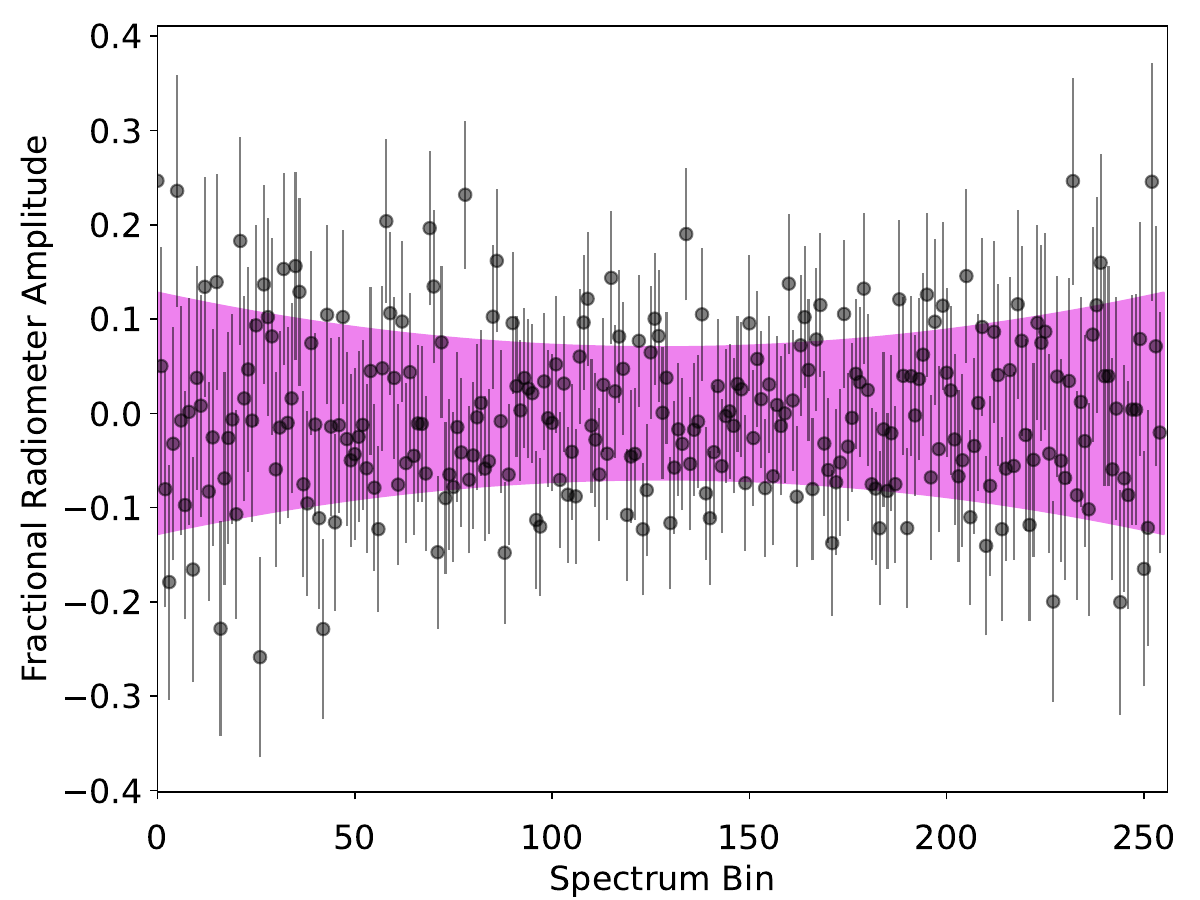}
\caption{Sample simulated normalized power spectra over mode's fitted Lorentzian transfer function with point-like errors (black, $Q_L = 30,000$) overlayed with modulated Johnson-distribution width (magenta).}
\label{fig:poweroverGamma}
\end{figure}

Now consider how each of the terms in Eqn.~\ref{eqn:chi2} are calculated. The first term involves only the scan data and need be calculated only once for each scan regardless of the hypothesized axion mass. Its computation scales as the number of bins in the scan. The other two terms of the test statistic contain factors of the transmitted axion lineshape into the scan and are refactored for more efficient calculation over multiple axion masses. For both terms, the Lorentzian cavity transmission function $T_{\nu_o}$ is factored out of the emitted axion power $P_{emitted} = T_{\nu_o} P_a$ and is instead used to modulate the noise error estimate $\sigma_w = \sigma/ T_{\nu_o}$ and the power spectrum $P_w = P_{emitted}/T_{\nu_o}$ so that they now vary over the spectrum, see Fig.~\ref{fig:poweroverGamma}. The bandwidth of each digitized power spectrum is on the order of the Q-width of the TM$_{010}$-like mode, less than $1:10^4$ of the central frequency. It is by this same fraction that the supposed axion line shape will change its width across mass tests spanned by a single scan. To reduce the computational costs of generating a unique axion line shape for each probed mass, the line shape is generated once for the scan's central frequency ($p_{a,i} = \int_{b_i} d \nu f(\nu_a - \nu)$) and is approximated as translationally invariant over the scan. The factor of frequency in the axion power equation is also held constant at the scan's center point. The second term of Eqn.~\ref{eqn:chi2} that computes the inner product of the power spectrum and line shape can now be phrased as a discretized convolution. Given proper zero-padding to the line shape and power spectrum, the intersection can be computed over the whole scan using the (discretized) convolution theorem
\begin{equation}
     p_{a} * \left( \frac{T_{\nu_0} \delta_w}{\sigma^2_{P}} \right) = \mathcal{F}^{-1} \left( \mathcal{F} \left( p_{a} \right) \cdot \mathcal{F} \left( \frac{T_{\nu_0} \delta_w}{\sigma^2_{P}} \right) \right),
\end{equation}
where the Fourier transform is discrete and implemented using the fast Fourier algorithm.

\begin{figure}
\centering
\includegraphics[width=\linewidth]{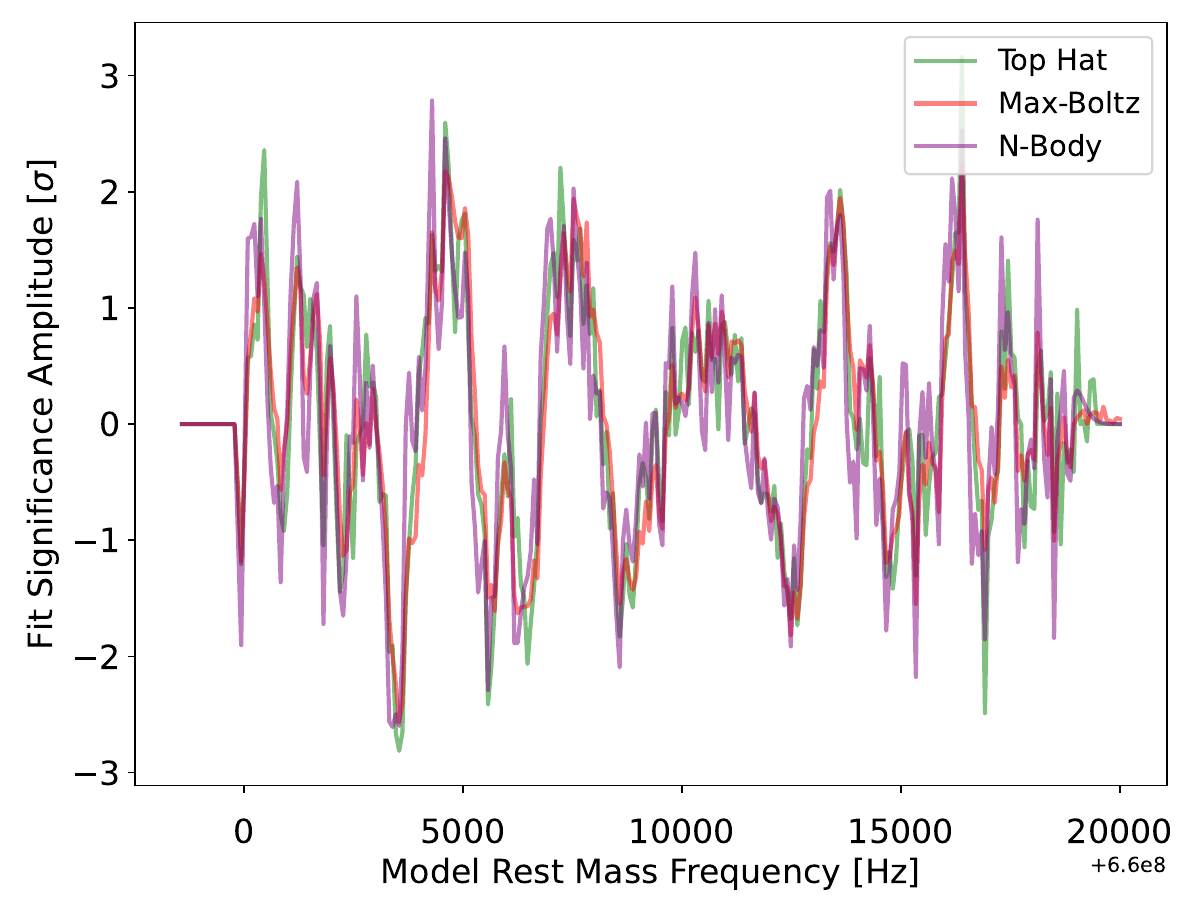}
\caption{Fit significance $\Sigma_A$ using the sample scan from Fig.~\ref{fig:poweroverGamma} for each line shape model used in the live and off-line analyses. Different line shape models are discussed in Section~\ref{sec:Signal}.}
\label{fig:significance}
\end{figure}

The third term of Eqn.~\ref{eqn:chi2} is the overlap of the squared axion line shape with the power spectrum's modulated error, and can also be phrased as a discrete convolution, using similar zero-padding as the second term,
\begin{equation}
    p^2_{a} * \left( \frac{T_{\nu_0}^2}{\sigma^2_{P}} \right) = \mathcal{F}^{-1} \left( \mathcal{F} \left( p^2_{a} \right) \cdot \mathcal{F} \left( \frac{T_{\nu_0}^2 }{\sigma^2_{P}} \right) \right) .  
\end{equation}
Now one can compute each of the statistics in the previous sub-section to compute quantities such as the fit significance of axion hypotheses, Fig.~\ref{fig:significance}. This use of convolutions to evaluate the data's senstivity to a given axion signal shape allows one to think of the search technique as an optimal filter.

\subsection{Grand Spectrum}

This sub-section integrates the statistics of single scan analyses into a grand spectrum of tests. This will be accomplished by deriving the arithmetic rules for the combination of single scan statistics for each figure of interest, then applying them to the set of viable scans.

Recall that the independence of each measurement allows the total likelihood function of Eqn.~\ref{totallikelihood} statistic to be decomposed as a product of single scan likelihoods. The total $\chi^2$ statistic therefore can also be decomposed into a sum of single scans
\begin{equation}
    \chi^2_{tot} = \sum_{s \in N_s} \chi_s^2
\end{equation}
where $N_s$ is the set of viable scans. Further, the decomposition of the statistic in powers of the axion signal strength also factorize to
\begin{align}
    \chi^2_{tot} &= \sum_{s \in N_s} \sum_{i_s \in s} \frac{\delta_{w, i_s}^2}{2 \sigma^2_{P_{i_s}}} - 2 A P_{DFSZ} \sum_{s \in N_s} \sum_{i_s \in s} \frac{T_{\nu_0,i_s} p_{a, i_s} \delta_{w, i_s}}{2 \sigma^2_{P_{i_s}}} \nonumber \\
    & + A^2 P_{DFSZ}^2  \sum_{s \in N_s} \sum_{i_s \in s} \frac{T_{\nu_0,i_s}^2 p_{a, i_s}^2 }{2 \sigma^2_{P_{i_s}}},
\end{align}
where the $i_s$ index runs over the bins of scan $s$. Recall that we have organized all the tests according to prescribed axion rest masses and not bin frequency, minimizing the complications from scans with mismatched ranges and bin boundaries.

The expectation and uncertainty statistics of interest to the grand spectrum are then seen to have the following addition rules
\begin{align}
    E(A)_{tot} &= \sum_{s \in N_s} h_s E(A_s), \\
    \sigma(A)_{tot} &= \frac{1}{\sqrt{\sum_{s \in N_s}\frac{1}{\sigma_{A_s}^2}}},
\end{align}
where $h_s$ are the expectation value weights given by
\begin{equation}
    h_s = \frac{\chi^2_{A^0, s}}{\sum_{k \in N_k} \chi^2_{A^0, k}}.
\end{equation}
One can also subtract scans from the set by reversing the sign of the arithmetic operation. These are used to identify candidates, and ultimately provide the exclusion limits in Sections~\ref{sec:Candidates} and \ref{sec:Limits}.

\section{Axion Signal Models}
\label{sec:Signal}

An axion signal emitted from the cavity is stimulated from the oscillations in the axion field passing through during the integration phase of a data-taking cycle. This section presents the models used to emulate the signal generated by the ambient axion dark-matter distribution. This analysis takes a similar approach to that of \citet{2018PhRvD..97l3006F} for modeling the response from a classical axion distribution, where the net axion field passing through the cavity during the observation is taken as a superposition of plane waves, which is considered accurate for a classical field on distances much shorter than the curvatures of an axion's orbit. The axion field inside the cavity is nearly homogeneous as its extent is on the order of the axion's Compton length, much less than the de Broglie length that sets the spatial fluctuations. The total axion field in the cavity is then 
\begin{equation}
    a(t) = \frac{\sqrt{2 \rho_{DM}/ N_a}}{m_a}\sum_i^{N_a} \cos \left( E_i t + \phi_i \right)
\end{equation}
where $\rho_{DM}$ is the local dark-matter density, $N_a = \mathcal{O} (V\rho_{DM}/ m_a)$ is the number of axions passing through the cavity at any one time, $E_i$ is the energy of the i-th axion as measured from the cavity's rest frame, and $\phi_i$ is the phase of the wave at the start of observation ($t=0$). Most axions passing through the cavity will be bound to the Milky Way halo, having speed less than the escape speed of $ \approx 560$~km/s relative to the galactic center \citep{Monari2018}. The motion of the cavity, also in orbit, has been measured to be of speed $230 \pm 5$~km/s \citep{Turner1990} co-rotating with spin of the Milky Way. The motion of the majority of axions is therefore of the order $\sim 10^{-3}$~c and the energy of each axion is well approximated by the lowest order kinematic expansion of the energy $E = m_a + m_a \Delta v^2/2 + O(\Delta v^4/c^4)$ where $\Delta v$ is the relative speed of the axion to the cavity.

The modulus squared of the axion field produces the power spectrum of the axion field as used in Eqn.~\ref{eqn:axionpwr} is then found to be
\begin{equation}
    |a(\nu)|^2 = \frac{2 \rho_{DM}}{m_a^2} f_{DM} \left(\delta \nu \right)
\end{equation}
where $\delta \nu = \nu - \nu_0$ is the frequency relative to the rest mass, and $f_{DM}(\delta \nu)$ is the local frequency distribution function of the axion halo. Several distribution functions were used in the search for axions during Run 1A, which are detailed in the following subsections.

\subsection{Top-Hat Model (Live Analysis)}

The first and simplest distribution model is a top-hat function the width of seven MR bins ($\sim 700$~Hz). This model roughly reproduces the width of the local axion distribution, which at $650$~MHz is expected to have an overall width in proportion to the expected virialaized speed squared $w \approx 650 \text{~MHz} \left< v^2 \right>/2 c^2 \sim 700$~Hz, and is the most robust of the models to detect power excesses at the expected width. The top-hat model is used only during the live analysis that occurs in parallel to data-taking operations, informing on the overall health of data and identifying axion candidates.

\subsection{Isotropic Isothermal Sphere}

The standard halo model (SHM) distribution is the most common shape used by axion searches and direct dark-matter searches in general. The SHM is based on the assumption that the MW halo is given by a thermalized pressure-less self-gravitating sphere of particles. More specifically, we use the truncated isothermal sphere model \citep{BT2008}, which is constructed with finite mass and has a cutoff at the halo escape speed.

In the frame of the galactic center, the velocity distribution at the solar radius takes on the near-Maxwell-Boltzmann form
\begin{align}
&f_v(\vec{v})  \propto
&\left\{
	\begin{array}{ll}
		4 \pi (\frac{1}{\sigma^2 \pi })^{\frac{3}{2}} v^2 e^{-\frac{v^2}{\sigma^2}}  & 0 \le |v| \le 560~km/s \\
		0 & \mbox{(otherwise) },
	\end{array}
\right. \label{eqn:MaxBoltz}
\end{align}
where $\sigma$ is the dark matter velocity dispersion. The approximation of a full Maxwell-Boltzmann distribution is adequate for the MR analysis and retains an analytic form when boosted from the galactic frame by $\vec{v_{lab}}$ into the cavity frame
\begin{equation} \label{equ: SHM}
f_v(\vec{v}) \approx \frac{2 \hbar c^2}{\sqrt{(2 \pi \sigma^2) M_{rest} v_{lab}}} \times \sinh{ \left(\frac{\vec{v} \cdot \vec{v_{lab}}}{2 \sigma^2}\right) } \times e^{\left(\frac{-(v^2 + v_{lab}^2)}{2 \sigma^2}\right)}
\end{equation}
where $\vec{v}_{lab}$ is the velocity of the lab relative to the galactic center. The motion of the lab during Run 1A was set to the orbital velocity of the sun around the galactic center. One could also incorporate the shifting motions of the Earth orbiting about the Sun and the Earth's spin, however these changes would only impact the overall width of the signal by less than 10\% \cite{Brubaker:2017rna}, and therefore were ignored during the analysis.

\begin{figure}
  \centering
  \includegraphics[width=\linewidth]{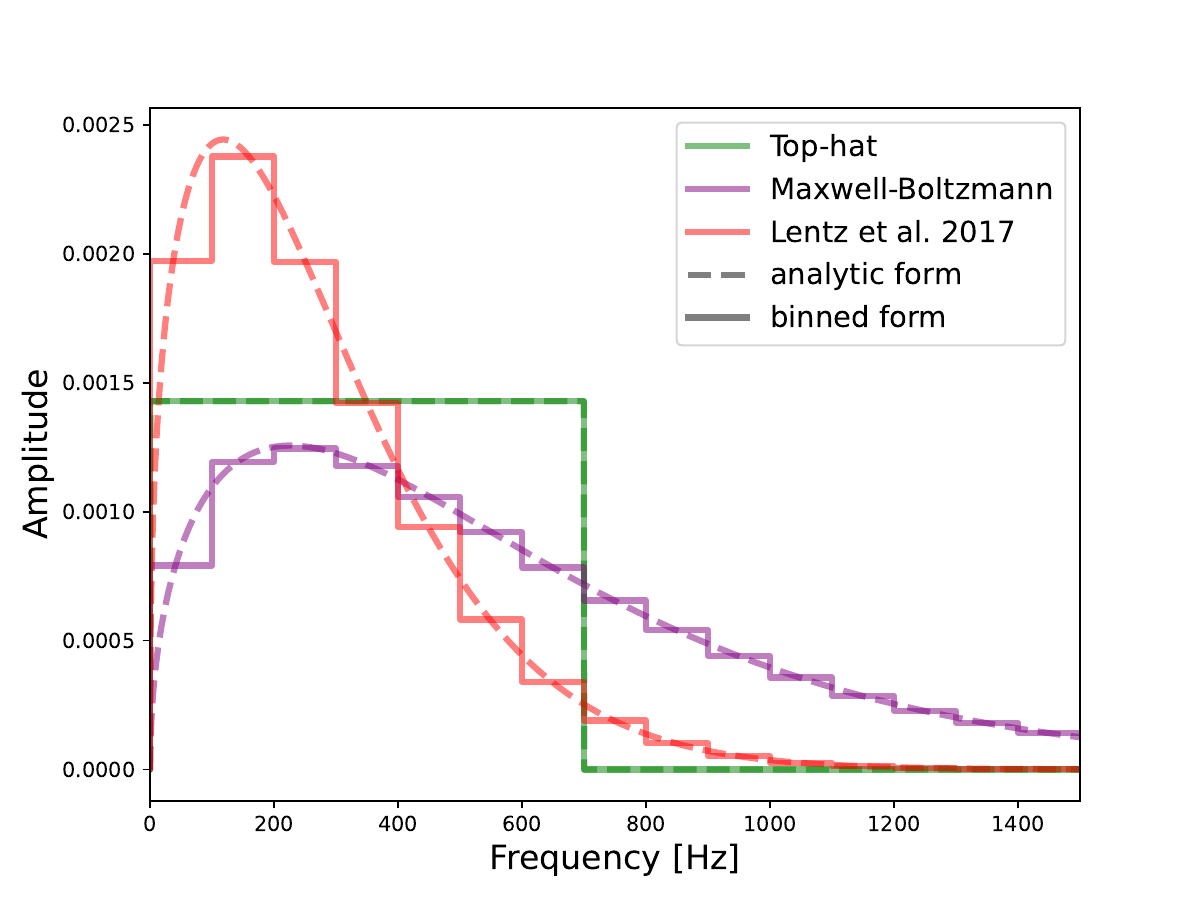}
  \caption{Sample unit-normalized energy distribution functions for an axion dark-matter candidate with rest mass corresponding to 650~MHz. Each line shape used in Run 1A is represented both in its smooth form (dashed) and as a binned sample (solid) where a bin edge is aligned with the axion rest mass. The width of the top-hat line shape was equivalent in width to seven MR bins and was not varied throughout the scan range. The widths of the Maxwell-Boltzmann and N-body inspired line shapes were varied on a scan-by-scan basis as described in Section~\ref{sec:Statistics}.}
  \label{fig:line shapes}
\end{figure}

\subsection{N-Body Model}

The third line shape used in the Run 1A analysis is rooted in the highly detailed Romulus25 cosmological simulation \cite{Lentz2017,Du2018,2017MNRAS.470.1121T}, where the halos of MW-like galaxies were found to be notably denser and have narrower line shape when sampled from the reference of a Sun-like orbit. The signal shape generated takes a Maxwell-Boltzmann-like form parameterized by three constants. The N-body inspired line shape takes the form
\begin{center}
\begin{equation} \label{NBODY}
f_\nu \propto \Bigg( \frac{(\nu - \nu_0)h}{m_a T} \Bigg)^{\alpha} e^{-\big( \frac{(\nu - \nu_0)h}{m_a T}\big)^\beta}
\end{equation}
\end{center}
where $\nu_0$ is the rest frame frequency, the exponent parameters are set to $\alpha = 0.36$, $\beta = 1.39$, and the distribution temperature is given by $T = 4.7 \times 10^{-7}$. This shape has a width that is 1.8 times narrower than the SHM line shape. The Romulus25 analysis also revealed a local expected DM density of $\rho_{DM} \approx 0.6$~GeV/cc, a notable increase from $0.45$~GeV/cc used in previous ADMX analyses.

\section{Candidate Handling}
\label{sec:Candidates}

\begin{figure}
  \centering
  \includegraphics[width=\linewidth]{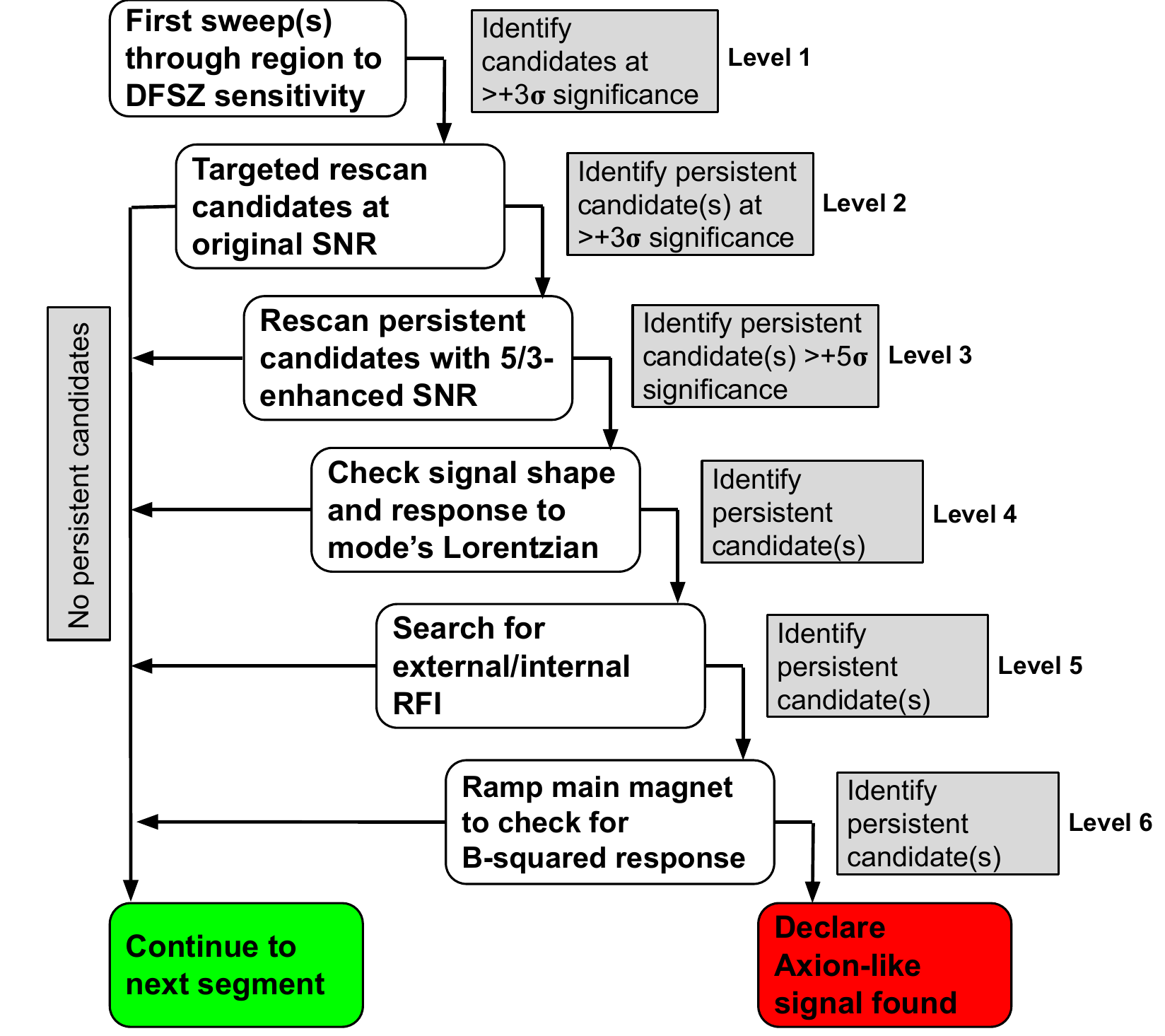}
  \caption{The decision tree for live candidate handling during Run 1A. The tree is entered into at the end of initial sensitivity scans of each 5-10~MHz section and traversed based on the identification and classification and candidate axion signals. Candidates are initially identified using the live analysis parameters.}
  \label{fig:candidatehandling}
\end{figure}

The processing and classification of candidate signals provides the means to claim detection of axion dark matter or to place limits on its mass and coupling. This section details the protocol used to identify, test, and re-test candidates through several filtering mechanisms unique to axion-photon conversion to robustly classify power excesses observed in the MR data. The decision protocol used for candidate handling in Run 1A is summarized in the decision tree of Fig.~\ref{fig:candidatehandling}. 

\subsection{Live Analysis}

A data analysis was run in parallel to data-taking operations, referred to as the live analysis, to inform the operators on the \textit{in situ} sensitivity of the data and to identify candidate axion signals. 

The live analysis operates on MR data under the chosen options to model the raw scan background using the six-order polynomial, and filters the prepared spectra using the top-hat signal model. These options were chosen for their low computational cost and robustness. Also, axion mass tests were made more sparsely to further reduce the computational cost, with separation of seven MR bin widths to reflect the size of the top-hat signal filter.

The significance statistic of Eqn.~\ref{eqn:significance} was used to search for candidate axion masses. The initial set of candidates were identified with a threshold of $+3 \sigma$, over which a particular mass test is considered to be a candidate. The expected number of candidates per 10~MHz segment in this mass range is $\approx
15$. During the candidate handling procedure, an excess of persistent and non-persistent candidates were found in segment 3. This excess of non-persistent candidates was found to be the result of inadequate background modeling by the six-order polynomial fit in the presence of significant background structure. Residual background structure larger than the noise floor were seen as power excesses (and deficits) by the analysis with enough regularity to skew the rate at which candidates were identified using a prescribed threshold. Also, a deficit of candidates was found in segment 1 by the live analysis. Possible systematic errors such as errant background subtraction or poorly fitted cavity Lorentzian were investigated but not found be the cause.

\begin{figure}
  \centering
  \includegraphics[width=\linewidth]{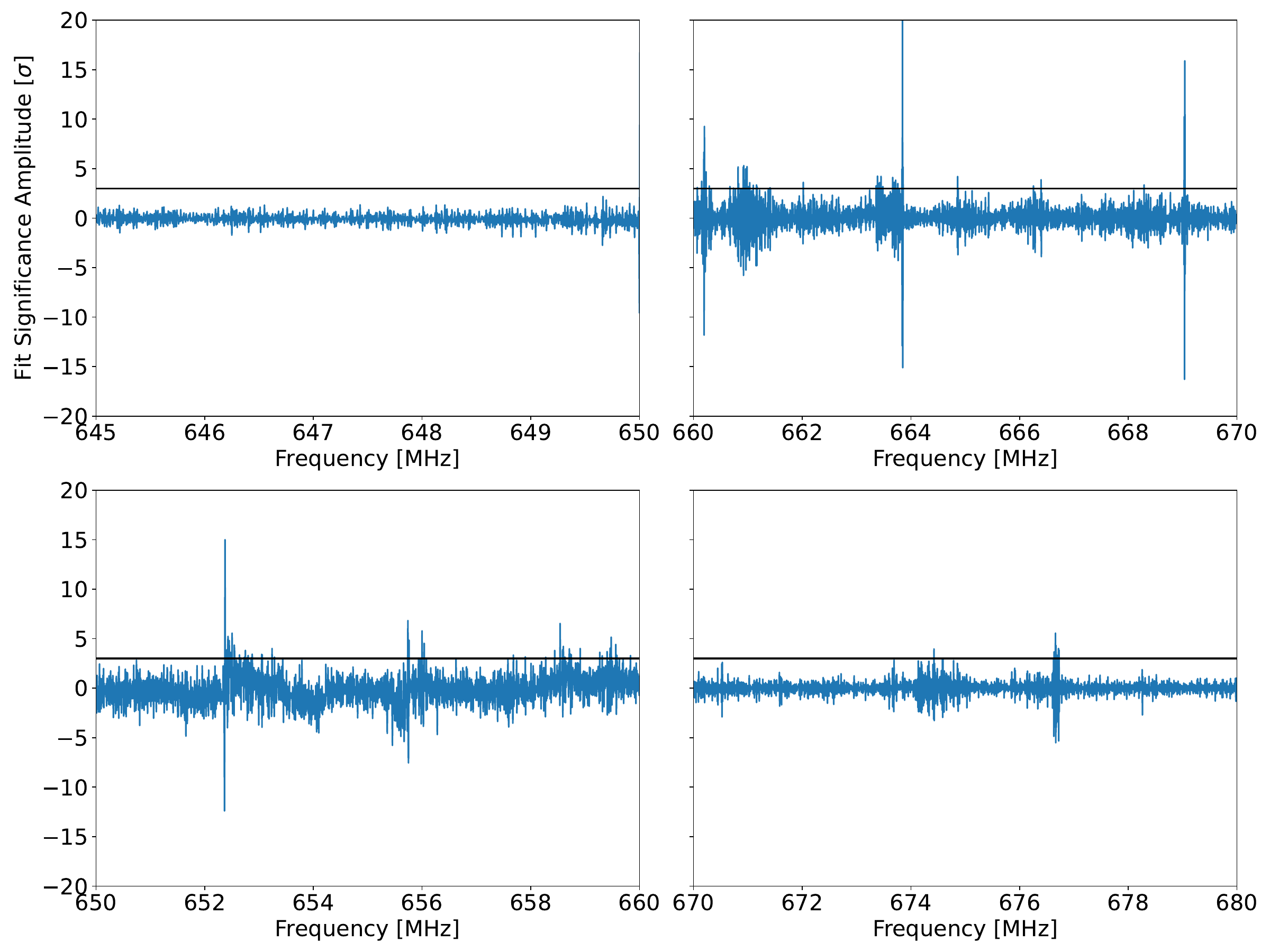}
  \caption{Candidate search results (Level 1) for each segment. Candidate status was determined from the significance measure $\Sigma_A$ using a $+3 \sigma$ threshold. The number of candidates observed above the threshold can be found in Table~\ref{table:candidates}, and overall were seen at a significantly higher rate than expected from purely thermal fluctuations. The top-hat shape was used to determine candidate status via the live analysis run in parallel with data taking operations. The Maxwell-Boltzmann and N-body line shapes were implemented in the offline analysis and found no additional candidates.}
  \label{fig:candidate selection}
\end{figure}

\subsection{Candidate Procedure}

Once candidates masses in a segment are identified, a targeted re-scan of the immediate area around each candidate frequency is conducted to the same sensitivity as the initial observations. The re-scan data was again analysed using the live analysis procedure to test for persistence using the same $+3 \sigma$ threshold. Candidates that did not persist were classified as random thermal variations. Persistent candidates were again rescanned, but to an SNR 5/3 that of the initial scans, corresponding to a slowdown in scan speed of 9/25. The re-scans are analysed for persistence using the same live analysis procedure with an updated threshold of $5 \sigma$. 
Candidate masses that do not pass this threshold are classified as random thermal variations and removed from further consideration. No thermal candidates are expected to pass this second rescan stage. 

Remaining candidates are put through further persistence tests that also test for axion and dark-matter specific qualities. The first process is a further rescan of the candidate, again at slowed rate, and a test not only of the signal's persistence over the rescan but also of the signal's response to the Lorentzian envelope of the coupled mode $T_{\nu_0} \times |a|^2$, which confirms that the cavity is the origin of the power excess. The test is performed using the live analysis procedure, checking for significance that scales like $1/T_{\nu_0}$. Persistent candidates are also tested at this point for a shape that matches either a Maxwellian distribution such as described in the previous section or a non-thermalized shape with dominant fine structure such as those proposed in \citet{Banik2016}. This is performed both with the MR data as well as the full time series, mixed to resolution at the Hz level. Given the large range of proposed signals, few candidates are ruled out by this step unless they are un-physically wide.

The penultimate test in the candidate protocol is an RFI search at the remaining candidate frequencies. Using an exterior antenna and signal analyser, ambient signals are searched for about the ADMX insert and DAQ. Candidates found to have a counterpart external RF signal at the same frequency and shape, or at intermediate frequencies of the receiver, are classified as RFI.

The final test in the Run 1A candidate handling protocol is a ramping of the main magnet while taking data about a single persistent candidate. The data is then analyzed for a $\propto B^2$ response in the emitted signal power particular to the inverse-Primakoff process. A signal exhibiting this response is categorized as a robust candidate for axion dark matter. Table~\ref{table:candidates} shows the number of candidates for each segment at each stage of the decision tree. No candidates survived past the fourth round.

\begin{table}[h!]
\centering
\begin{tabular}{ |p{2cm}|p{1cm}|p{1cm}|p{1cm}|p{1cm}|p{1cm}|p{1cm}|  }
 \hline
 Segment & Level 1 & Level 2 & Level 3 & Level 4 & Level 5 & Level 6 \\
 \hline
 645-650 & 0 &  0 & 0 & 0 & 0 & 0\\
 650-660 & 14 & 1 & 0 & 0 & 0 & 0\\
 660-670 & 34 & 5 & 3 & 2 & 0 & 0\\
 670-680 & 17 & 0 & 0 & 0 & 0 & 0\\
 \hline
\end{tabular}
\caption{Number of candidates present at each stage of the candidate handling decision tree. The identified RFI in the region $660.16-660.27$~MHz was unfortunately found to change its intensity and shape with time and was unable to be subtracted from the power spectrum. Axion tests overlapping with these signals were omitted when calculating the final exclusion limits.}
\label{table:candidates}
\end{table}

\subsection{Synthetic Axions}

Run 1A saw the beginnings of a candidate injection system where artificial signals were injected into the power spectrum, using both hardware and software methods, in order to test instrument sensitivity.

Hardware axions were injected into the cavity via the weak port through with the Synthetic Axion Generator (SAG). The structure of the SAG and its placement in the RF system has been presented in \citet{Khatiwada2020}.

Injection frequencies were set prior to the scanning of each segment. Data from data-taking cycles with an injection were flagged and assessed independent of the main axion search. Blind injection did not occur until Run 1B \cite{Khatiwada2020, Bartram2021}. A second complete data-taking cycle at the same tuning rod configuration was taken immediately following an injection so as to not mask the data in the vicinity of the synthetic's central frequency. Only the second scan was used in the axion search analysis presented here.

\begin{figure}
\begin{centering}
\includegraphics[width=0.5\textwidth]{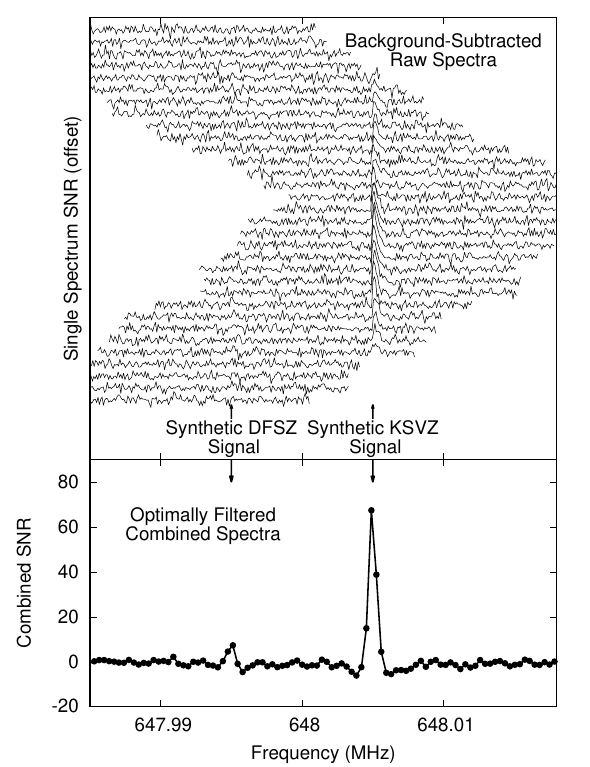}
\caption{(Upper) Series of background-subtracted single scans with synthetic axion signals with the N-body inspired signal shape \cite{Lentz2017}, one at KSVZ coupling and one at DFSZ coupling.  The KSVZ signal is easily visible in these individual spectra; the DFSZ signal, being a factor of 7 smaller, is not. (Lower) Same data after the individual scans have been optimally filtered and combined.  Both KSVZ and DFSZ signals are visible with high SNR. Figure first appeared in \citet{Du2018}. \label{fig:waterfall}}
\end{centering}
\end{figure}

Software injections of axion signals were also used to test the performance of the analysis. Simulated axion signals were imposed directly onto the raw power spectra in the forms of the Maxwellian and N-body inspired lineshapes, modulated by the fitted Lorentzian response of the cavity. We injected 25,000 software-simulated signals into the data set with couplings varying between DFSZ and 10 times KSVZ, ran through the analysis process, and evaluated the resulting candidate power to determine the systematic uncertainty associated with the background subtraction. Figure~\ref{fig:waterfall} shows the effect of the injected signals in both the background-subtracted spectra and the final filtered and combined spectrum.

\section{Exclusion Limits}
\label{sec:Limits}

The previous section showed how the candidate procedure identified and classified candidates in the grand axion search analysis. Each of these greater-than-$+ 3 \sigma$ candidates were classified as non-axion in origin, therefore making the observations consistent at that level with the null-signal hypothesis under the live analysis top-hat line shape. Analyses using the Maxwellian and the N-body axion lineshapes were performed after the run's data-taking operations. They both found no new candidates at the same threshold.

\begin{figure*}
\centering
\includegraphics[width=\textwidth]{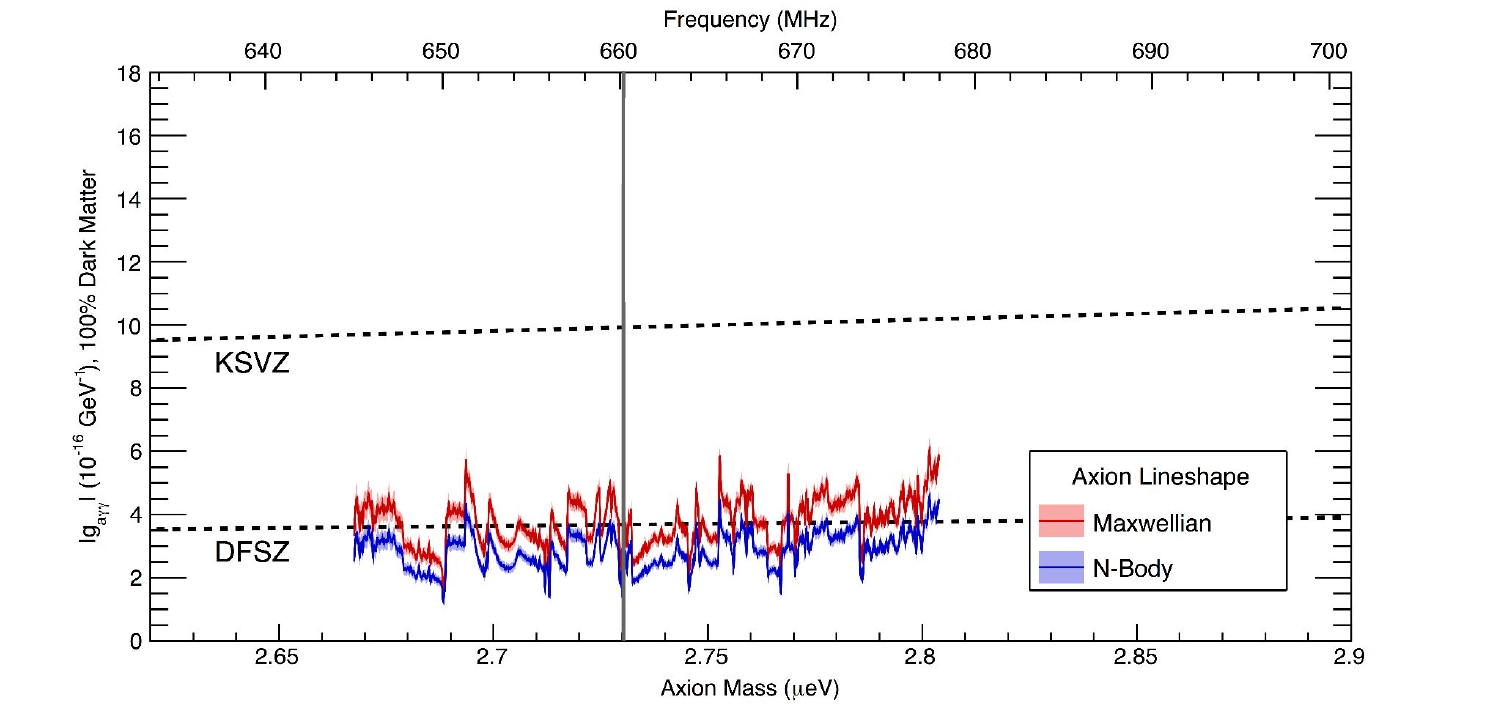}
\caption{ The 90\% upper confidence excluded region of axion mass and photon coupling $g_{a \gamma \gamma}$, modified from figure as seen in \citet{Du2018}. The red line indicates the limit on axion-photon coupling with the boosted Maxwell-Boltzmann line shape from the isothermal halo model \citep{Turner1990}, while the blue line indicates the limit with the N-body inspired signal \citep{Lentz2017}. Colored shaded regions indicate the systematic uncertainty range. The region 660.16 to 660.27~MHz marked by the gray bar was vetoed due to interference as described in the text. The inset shows the results in the context of other haloscope searches.}
\label{fig:Run1Alimits}
\end{figure*}

Given the null outcome, upper limits can then be formed on the axion power $P_a$, and therefore coupling $g_{a \gamma \gamma}$ under the priors presented in Sec.~\ref{sec:Statistics} for each axion mass tested. The one-sided 90\% limits on the power-derived coupling are shown in Fig.~\ref{fig:Run1Alimits} for both the SHM and N-body line shapes. 

\section{Summary and Conclusions}
\label{sec:Conclusions}

This paper elaborates on the background and methods used by the ADMX collaboration to gather its Run 1A data and perform the axion dark matter search first reported on in \citet{Du2018}, which produced the first axion-photon coupling limits at DFSZ sensitivity. The ADMX apparatus was reviewed, with special concentration on the microwave cavity, receiver, and the magnet system that enabled resonant conversion of an ambient coherent axion field. Data-taking operations were explicated from their global structure down to the individual measurements of a single data-taking cycle. Interpretations of the raw data and its preparation for axion search were then presented, including the classification of the cavity and receiver transmission properties, noise temperatures, and structure of the receiver chain emitted noise power in the limit of many photon emissions per observation. The statistical basis for the axion search was then presented, both at the level of a single observation and the level of the wider Run 1A data set, and co-added into a grand spectrum spanning the observation range. The space of lineshapes for the local axion distribution and their usage were presented for both the live analysis and the off-line analyses. The axion signal candidate handling procedure was outlined, with candidates observed in the Run 1A data being catalogued either as random thermal fluctuations or interference external to the insert. No axion candidates persisted through the candidate-search process, making the data consistent with the null axion hypothesis and allowing detection limits to be formed over the observed range, with the exception of a set of RFI signals that could not be reliably removed from the data due to their strong time-dependent properties.

Several updates and improvements have been made to the instrumentation and analysis that have since been used for Run 1B \citep{Braine2020,Bartram2021} and Run 1C still in progress \citep{Nitta2021}. Areas highlighted in the Run 1A analysis are as follows: an improved modeling of the receiver structure during live analysis to a Pad\`e filter from the six-order polynomial, which drastically reduces the number of candidates found in the initial search, bringing the number to near that expected from purely thermal contributions; synthetic axion injection, which was tested but not thoroughly integrated in Run 1A, has been fully incorporated into the data acquisition process using both hardware injected blinded and un-blinded candidates and software candidates that have been used to good effect to improve confidence in a search's sensitivity to the axion; and last to be mentioned here, the MSA quantum-limited amplifier of Run 1A has been replaced by JPAs (Josephson Parametric Amplifiers) in Runs 1B and 1C and have proven to be more stable and tunable.

Lastly, the possibility of fine structure existing in the local axion energy distribution presents a distinct opportunity to improve the observation's sensitivity and has motivated the collaboration to store the full time series of each observation. The resolution of this data extends to the tens of milli-hertz, the optimal level between the finest motivated fine structure features \citep{Banik2016} and smearing due to orbital modulation from the Earth's spin and motion around the solar system center of mass. This high resolution data is significantly more sensitive to such fine structures. Therefore, as ADMX brings in more data at DFSZ sensitivity or better, searches for fine structure will play an increasingly important role in furthering the discovery potential for axion dark matter at increasingly smaller couplings or proportions of the total dark-matter density. 

\section{Acknowledgements}

This work was supported by the U.S. Department of Energy through Grants No. DE-SC0009800, No. DESC0009723, No. DE-SC0010296, No. DE-SC0010280, No. DE-SC0011665, No. DEFG02-97ER41029, No. DEFG02-96ER40956, No. DEAC52-07NA27344, No. DEC03-76SF00098, No. DE-SC0017987, and No. DE-SC0022148. Fermilab is a U.S. Department of Energy, Office of Science, HEP User Facility. Fermilab is managed by Fermi Research Alliance, LLC (FRA), acting under Contract No. DE-AC02-07CH11359. Additional support was provided by the Heising-Simons Foundation and by the Lawrence Livermore National Laboratory and Pacific Northwest National Laboratory LDRD offices. LLNL Release No. LLNL-JRNL-848336. Chelsea Bartram acknowledges support from the Panofsky Fellowship at SLAC.

\newpage

\bibliography{bibliography.bib}

\end{document}